\documentclass[12pt, draftclsnofoot, journal, letterpaper, onecolumn]{IEEEtran}

\makeatletter
\def\ps@headings{%
\def\@oddhead{\mbox{}\scriptsize\rightmark \hfil \thepage}%
\def\@evenhead{\scriptsize\thepage \hfil \leftmark\mbox{}}%
\def\@oddfoot{}%
\def\@evenfoot{}}
\makeatother 
\IEEEoverridecommandlockouts
\usepackage{bbm}
\usepackage{amsfonts}
\usepackage[dvips]{graphicx}
\usepackage{times}
\usepackage{cite}
\usepackage{amsmath}
\usepackage{array}
\usepackage{amssymb}

\newcommand{\bs}{\boldsymbol}

\usepackage{stfloats}
\usepackage{graphicx}
\usepackage{footnote}
\usepackage{booktabs}
\usepackage{array}
\usepackage{algorithm}
\usepackage{subeqnarray}
\usepackage{cases}
\usepackage{threeparttable}
\usepackage{color}
\usepackage{hyperref}
\usepackage{epstopdf}
\usepackage{algpseudocode}
\usepackage{bm}
\usepackage{multirow}
\usepackage[labelformat=simple]{subcaption}
\usepackage{adjustbox}
\usepackage{enumitem}

\newtheorem{theorem}{\bf Theorem}

\newtheorem{lemma}{\bf Lemma}
\newtheorem{definition}{\bf Definition}

\newtheorem{corollary}{\bf Corollary}

\allowdisplaybreaks

\begin{document}

\title{\huge Joint Status Sampling and Updating for Minimizing Age of Information in the Internet of Things}

\author{Bo~Zhou,~\IEEEmembership{Member,~IEEE} and Walid~Saad,~\IEEEmembership{Fellow,~IEEE}
\thanks{This research was supported by the U.S. National Science Foundation under Grants IIS-1633363, CNS-1836802, and CNS-1460316.
A preliminary version of this work has been presented at IEEE GLOBECOM 2018\cite{gc18}.
B.~Zhou and W.~Saad are with Wireless@VT, Bradley Department of Electrical and Computer Engineering, Virginia Tech, Blacksburg, VA 24061, USA.
Email: \{ecebo, walids\}@vt.edu.}
}

\maketitle
\vspace{-1.5cm}

\begin{abstract}
The effective operation of time-critical Internet of things (IoT) applications requires real-time reporting of fresh status information of underlying physical processes.
In this paper, a real-time IoT monitoring system is considered, in which the IoT devices sample a physical process with a sampling cost and send the status packet to a given destination with an updating cost. This joint status sampling and updating process is designed to minimize the average age of information (AoI) at the destination node under an average energy cost constraint at each device. This stochastic problem is formulated as an infinite horizon average cost constrained Markov decision process (CMDP) and transformed into an unconstrained Markov decision process (MDP) using a Lagrangian method. For the single IoT device case, the optimal policy for the CMDP is shown to be a randomized mixture of two deterministic policies for the unconstrained MDP, which is of threshold type.
This reveals a fundamental tradeoff between the average AoI at the destination and the sampling and updating costs.
 Then, a structure-aware optimal algorithm to obtain the optimal policy of the CMDP is proposed and the impact of the wireless channel dynamics is studied while demonstrating that channels having a larger mean
channel gain and less scattering can achieve better AoI performance.
For the case of multiple IoT devices, a low-complexity \textcolor{black}{semi-distributed} suboptimal policy is proposed with the updating control at the destination and the sampling control at each IoT device.
Then, an online learning algorithm is developed to obtain this policy, which can be implemented at each IoT device and requires only the local knowledge and small signaling from the destination.
The proposed learning algorithm is shown to converge almost surely to the suboptimal policy.
Simulation results show the structural properties of the optimal policy for the single IoT device case; and show that the proposed  policy for  multiple IoT devices outperforms a zero-wait baseline policy, with average AoI reductions reaching up to 33\%.

\end{abstract}


\section{Introduction}

% Recently, with the rapid proliferation of the Internet of Thing (IoT) devices, the demand for timely information has been increasing in many real-world systems. Such systems range from information updates, e.g.,  , in social networks , to status updates in real-time monitoring and control systems, such as environment monitoring in sensor networks, channel state information estimation in cellular networks, and vehicle tracking in smart transportation systems.

With the rapid proliferation of the Internet of Thing (IoT) devices, delivering timely status information of the underlying physical processes has become increasingly critical for many real-world IoT and cyber-physical system applications\cite{ATZORI20102787,7412759}, such as environment monitoring in sensor networks and vehicle tracking in smart transportation systems.
Given the criticality of IoT applications, it is imperative to maintain the status information of the physical process at the destination nodes as fresh as possible, for effective monitoring and control.

To quantify the freshness of the status information of the physical process, the concept of \emph{age of information} (AoI) has been proposed as a key performance metric\cite{6195689} that quantifies the time elapsed since the generation of the most recent IoT device status packet received at a given destination.
In contrast to conventional delay metrics, which measure the time interval between the generation and the delivery of each individual packet, the AoI considers the packet delay and the generation time of each packet, and, hence, characterizes the freshness of the status information from the perspective of the destination.
% Different from conventional delay metrics, which represent the freshness of the status information \textcolor{black}{with respect to} each individual packet \textcolor{black}{(i.e., the time interval between the generation and the delivery of each packet)}, the AoI characterizes the freshness of the  status information from the perspective of the destination\cite{kadota2018scheduling}.
Therefore, optimizing the AoI in an IoT would lead to distinctively different system designs from those used for conventional delay optimization.
\textcolor{black}{For example, it has been shown that the last-come-first-served (LCFS) principle achieves a lower AoI than the conventional first-come-first-served (FCFS) principle\cite{6310931}.}

\textcolor{black}{The problem of minimizing the AoI} has attracted significant recent attention \cite{6195689,6310931,hsu2017scheduling,kadota2018scheduling,HsuISIT,jiang2018can,jiang2018timely,8000687,bedewy2018age,8123937,bacinoglu2017scheduling,feng2018minimizing,wcncage,sun2017remote}.
Generally, these works can be classified into two broad groups based on the model of the generation process of the status packets.
The first group \cite{6195689,6310931,kadota2018scheduling,hsu2017scheduling,HsuISIT,jiang2018can,jiang2018timely} models the generation process of the status packets as a queueing system in which the status packets arrive at the source node stochastically and are queued before being forwarded to the destination. Queueing theory has also been used  to analyze and optimize the average AoI for FCFS \cite{6195689} and LCFS systems \cite{6310931}.
% In \cite{kadota2018scheduling}, the authors study the optimal scheduling in a wireless broadcast networks
\textcolor{black}{The works in \cite{hsu2017scheduling,kadota2018scheduling,HsuISIT} propose scheduling schemes that seek to minimize the average AoI in wireless broadcast networks.
In \cite{jiang2018can} and \cite{jiang2018timely}, the authors study the problem of AoI minimization in wireless multiaccess networks and propose decentralized scheduling policies with near-optimal performance.}
In the second group of works \cite{8000687,bedewy2018age,8123937,bacinoglu2017scheduling,feng2018minimizing,wcncage}, the status packets can be generated at any time by the source node.
 \textcolor{black}{The authors in \cite{8000687} and \cite{bedewy2018age} propose optimal updating policies to minimize the average AoI for status update systems, with a single source and multiple sources, respectively.
In \cite{8123937,bacinoglu2017scheduling,feng2018minimizing}, the authors propose optimal status updating schemes for an energy harvesting source to minimize the average AoI.}
The authors in \cite{wcncage} introduce an optimal status updating scheme  to minimize the average AoI under resource constraints.
 % \textcolor{black}{The work in \cite{xiao2018dynamic} studies the problem of AoI minimization in an adversarial setting.}
\textcolor{black}{Motivated by recent research on AoI, the authors in \cite{sun2017remote} study the remote estimation problem for a Wiener process and propose an optimal sampling policy to minimize the estimation error.}

In the existing literature, e.g., \cite{6195689,6310931,kadota2018scheduling,hsu2017scheduling,HsuISIT,jiang2018can,jiang2018timely,8000687,bedewy2018age,8123937,bacinoglu2017scheduling,feng2018minimizing,wcncage,sun2017remote}, the source node is usually required to perform simple monitoring tasks, such as reading a temperature sensor, and, hence, the cost for generating status packets is assumed to be negligible.
 % and the status packets can be transmitted to the destination instantly.
 However, next-generation IoT devices can now perform more complex tasks\footnote{One practical example is the Nest Cam IQ indoor security camera, which uses on-device vision processing to watch for motion, distinguish family members, and send alerts if someone is not recognized\cite{camera}.}, such as initial feature extraction and classification for computer vision applications, by using neural networks  and on-device artificial intelligence\cite{chen2017machine,teerapittayanon2017distributed}.
For such applications, generating the status update packets incurs energy cost for the IoT devices.
Moreover, compared to the status packets generated for simple monitoring tasks (e.g., a temperature reading), a generated status packet for sophisticated artificial intelligence tasks carries richer information on the underlying physical systems (e.g., objects detected in an image or video sequence).
Therefore, there will also incur some energy cost and \textcolor{black}{time delay} for transmitting those status packets with relatively large size to the destination node.
In presence of the energy cost  pertaining to the sampling and updating processes, a key open problem is to study how to intelligently sample the underlying physical systems and send status packets to the destination, in order to minimize the AoI.

% The main contribution of this paper is thus to characterize the optimal sampling and updating policy to minimize the average AoI at the destination under the average energy cost constraints of the IoT devices, by taking into account the energy cost and the time consumption for status sampling and updating. We formulate this stochastic optimization problem as an infinite horizon average cost constrained Markov decision process (CMDP) \cite{cmdp} and transform the CMDP into a parameterized unconstrained Markov decision process (MDP) using a Lagrangian method.
% We consider both the single and multiple IoT devices cases.
% For the case of a single IoT device, we characterize the structural properties of the optimal sampling and updating policy, propose a structural-aware optimal algorithm, and study the impacts of the wireless fading channels.
% For the case of multiple IoT devices, we develop an auction-based distributed low-complexity online algorithm to obtain a suboptimal sampling and updating policy, and show the almost surely convergence of the proposed algorithm.

The main contribution of this paper is, thus, to jointly design the status sampling and updating processes that can minimize the average AoI at the destination under an average energy cost constraint for each IoT device, by taking into account the energy cost  for generating and updating status packets.
 In particular, our key contributions include:
\begin{itemize}
% \item We formulate the stochastic optimization problem as an infinite horizon average cost constrained Markov decision process (CMDP) \cite{cmdp} and transform the CMDP into a parameterized unconstrained Markov decision process (MDP) using a Lagrangian method.
% Both the single and multiple IoT devices cases are investigated.

\item

  % $\bullet$
  For the single IoT device case, we formulate this stochastic control problem as an infinite horizon average cost constrained Markov decision process (CMDP)  and transform the CMDP into a parameterized unconstrained Markov decision process (MDP) using a Lagrangian method.
We show that the optimal policy for the CMDP is a randomized mixture of two deterministic policies for the unconstrained MDP.
By using the special properties of the AoI dynamics, we derive key properties of the value function for the unconstrained MDP. Based on these properties, we show that the optimal sampling and updating process for the unconstrained MDP is threshold-based with  the AoI state at the device and the AoI state at the destination.
% By using the value iteration algorithm \cite{bertsekas} and the special properties of the AoI dynamics, we derive key properties of the value function for the unconstrained MDP.
% Based on these properties, we show that the optimal sampling and updating process for the unconstrained MDP is threshold-based with respect to the AoI state at the device and the AoI state at the destination.
This reveals a fundamental tradeoff between the average AoI at the destination and the sampling and updating costs.
Then, we propose a structure-aware optimal algorithm to obtain the optimal policy for the CMDP.
We also study the influence of the wireless channel fading distribution on the optimal average AoI at the destination. By using the concept of stochastic dominance, we show that channels having a larger mean channel gain and less scattering can achieve better AoI performance.

% $\bullet$
\item
For the case of multiple IoT devices, \textcolor{black}{to obtain the optimal sampling and updating policy}, we also formulate a CMDP  and convert it to an unconstrained MDP.
We show that the optimal sampling and updating policy, which adapts to the AoI and channels states of all IoT devices, is a function of the Q-factors of the unconstrained MDP. To overcome the curse of dimensionality and to distribute the system's controls, we propose a low-complexity semi-distributed suboptimal policy by approximating the optimal Q-factors into the sum of per-device Q-factors, based on approximate dynamic programming. Then, we propose an online learning algorithm that allows each device to learn its per-device Q-factor, which requires only the  knowledge of the local AoI and channel states, as well as small signaling from the destination.
The proposed semi-distributed online learning algorithm is shown to converge almost surely to the proposed suboptimal policy.

% \item For the case of multiple IoT devices, to overcome the curse of dimensionality and to distribute the system's controls, we propose a distributed low-complexity  auction-based sampling and updating policy by approximating the optimal Q-factor functions into the sum of per-device Q-factor functions, based on approximate dynamic programming\cite{bertsekas}.
% Then, we propose a localized online learning algorithm for each IoT device to determine its per-device Q-factor and the associated Lagrange multiplier based on the real-time  local observations of the AoI and channel states. We prove that the proposed auction-based distributed online learning algorithm converges almost surely.

\item
% $\bullet$
 We provide extensive simulations to illustrate additional structural properties of the optimal policy for the single device case.
We show that the optimal thresholds for sampling and updating are non-decreasing with  the sampling cost and the updating cost, respectively, and the optimal action is also threshold-based with respect to the channel state.
For the case of multiple devices, numerical results show that the proposed semi-distributed policy outperforms a zero-wait baseline policy (i.e., sampling immediately after updating), with average AoI reductions  reaching up to $33\%$.
In summary, the derived results provide novel and holistic insights on the design of AoI-aware sampling and updating in practical IoT systems.
\end{itemize}

The rest of this paper is organized as follows.
In Section II, we present the single IoT device model and analyze its properties.
In Section III, we present the analysis for the case of multiple IoT devices using online learning.
\textcolor{black}{Section IV presents and analyzes numerical results.}
Finally, conclusions are drawn in \textcolor{black}{Section V}.

\section{Optimal Sampling and Updating Control for A Single IoT Device}

\subsection{System Model}

\begin{figure}[!t]
\begin{centering}
% \vspace{-0.6cm}
\includegraphics[scale=.6]{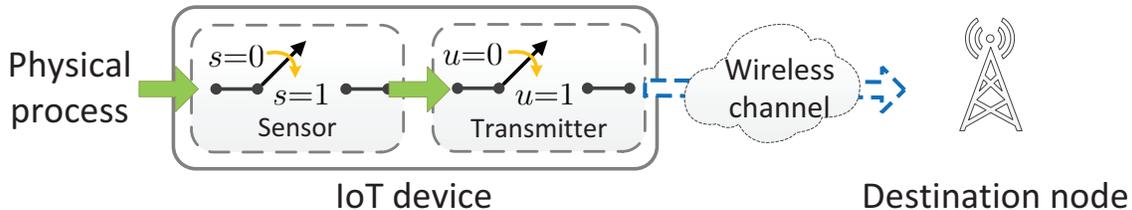}
 \caption{\textcolor{black}{Illustration of a real-time monitoring system with a single IoT device.}}\label{fig:system}
\end{centering}
% \vspace{-0.6cm}
\end{figure}

Consider a real-time IoT monitoring system composed of a single IoT device and a destination node (e.g., a base station or control center), as illustrated in Fig.~\ref{fig:system}. The IoT device encompasses a sensor which can monitor the real-time status of a physical process (referred to hereinafter as \emph{status sampling}) and a transmitter which can send status information packets to the destination through a wireless channel (referred to hereinafter as \emph{status updating}).
For the status sampling process, different from the existing literature where the device is usually assumed to perform simple sampling tasks \cite{6195689,6310931,kadota2018scheduling,hsu2017scheduling,HsuISIT,jiang2018can,jiang2018timely,8000687,bedewy2018age,8123937,bacinoglu2017scheduling,feng2018minimizing,wcncage,sun2017remote}, e.g., temperature and humidity monitoring, here, we consider that the IoT device can perform more sophisticated tasks, e.g., initial feature extraction and pre-classification using machine learning and neural network tools.
Hence, the time for status sampling and updating is not negligible and there will be some associated energy expenditures, which constrain the operation of the IoT device.

We consider a time-slotted system with unit slot length (without loss of generality) that is indexed by $t = 1, 2,\cdots$.
Let $h(t)\in\mathcal{H}$ be the channel state, representing the channel gain at slot $t$, where $\mathcal{H}$ is the finite channel state space.
We assume a block fading wireless channel over all time slots and we consider an i.i.d. channel state process $\{h(t)\}$ that is distributed according to a general distribution $p_{\mathcal{H}}(h)$.
Note that, the analytical framework and results can be readily extended to the Markovian fading channels.

\subsubsection{Monitoring Model}
In each slot, the IoT device must decide  whether to generate a status packet and whether to send to the status packet to the destination.
Let $s(t)\in\{0,1\}$ be the sampling action of the device at slot $t$, where $s(t)=1$ indicates that the device samples the physical process and generates a status packet at slot $t$, and $s(t)=0$, otherwise. We consider that a newly generated status packet will replace the older one  at the device, as the destination will not benefit from receiving an outdated status update.
This is similar to the LCFS principle explored in \cite{6310931}.
Let $C_s$ be the sampling cost for generating the status packet. This cost captures the computational cost needed for running some pre-classification algorithms using neural network models. We assume that the status sampling process takes one time slot.
Let $u(t)\in\{0,1\}$ be the update action of the device at slot $t$, where $u(t)=1$ indicates that the device sends the status packet to the destination at slot $t$ and $u(t)=0$, otherwise. The IoT device can only send the status packet available locally.
We denote by $C_u(h)$ the minimum transmission power required by the IoT device for successfully updating a status packet to the destination  within a slot when the channel state is $h$. Without loss of generality, we assume that $C_u(h)$ is decreasing with $h$.

Let $\bs{w}(t)\triangleq(s(t),u(t))\in\mathcal{W}\triangleq\{0,1\}\times\{0,1\}$ be the control action vector of the IoT device at  $t$. Then, the energy cost at the device associated with action $\bs{w}(t)$ is given by:
% \begin{equation}
$C(\bs{w}(t)) \triangleq s(t) C_s + u(t)C_u(h(t)).$
% \end{equation}

\subsubsection{Age of Information Model}
We adopt the AoI as the key performance metric to quantify the freshness of the status information packet \cite{6195689}. The AoI is essentially defined as the time elapsed since the generation of the last status update of the physical process. Let $A_r(t)$ be the AoI at the destination at the beginning of slot $t$. Then, we have
% \begin{equation}
% A_r(t) = t - \delta(t),
% \end{equation}
$A_r(t) = t - \delta(t)$,
where $\delta(t)$ is the time slot during which the most up-to-date status packet received by the destination was generated.
Note that, the IoT device can only send its currently available status packet to the destination. Thus, the AoI at the destination depends on the AoI at the device, i.e., the age of the status packet at the device.
Let $A_l(t)$ be the AoI at the device at the beginning of slot $t$.
The AoI at the device and the AoI at the destination are maintained by the device and can be implemented using counters.
Let $\hat{A}_l$ and $\hat{A}_r$ be, respectively, the upper limits of the corresponding counters for the AoI at the device and the AoI at the destination.
\textcolor{black}{We assume that $\hat{A}_l$ and $\hat{A}_r$ are finite.
 This is due to that, for time-critical IoT applications, it is not meaningful for the destination node to receive a status information with an infinite age. Such highly outdated status information will not be of any use to the system or underlying application.
Note that, the obtained results hold for arbitrarily finite $\hat{A}_l$ and $\hat{A}_r$, no matter how small or large $\hat{A}_l$ and $\hat{A}_r$ are.} 
% \footnote{This assumption guarantees that the AoI state space is finite which simplifies the analysis.}.
% This assumption guarantees that the AoI state space is finite, which would simplify the mathematical arguments \cite{cmdp}.
% \textcolor{black}{These finite values indicate that the status information is too stale to be of any use at the destination.} 
We denote by $\mathcal{A}_l\triangleq\{1,2,\cdots,\hat{A}_l\}$ and $\mathcal{A}_r\triangleq\{1,2,\cdots,\hat{A}_r\}$ the state space for the AoI at the device and the AoI at the destination, respectively.
We also define $\bs{A}(t)\triangleq(A_l(t),A_r(t))\in\mathcal{A}$ as the system AoI state at the beginning of slot $t$, where $\mathcal{A}\triangleq\mathcal{A}_l\times\mathcal{A}_r$ is the system AoI state space.

% This assumption is to guarantee that the request queue state space $\bm{\mathcal{Q}}$ is finite. According to [Puterman, Theorem 8.4.5], for unichain infinite horizon average cost MDPs with finite state and action spaces, and bounded costs, there always exists a stationary average optimal policy. Thus, based on this assumption, we would be able to design and analyse the optimal and suboptimal multicast scheduling policies. However, the obtained results for the finite state space cannot be straightforwardly extended to the infinite state space. For the infinite state space MDPs, much more rigorous mathematical analysis is usually required [Bertsekas, Chapter 4.6].
% Since our focus is to design the optimal multicast scheduling, to simply the arguments and avoid unnecessary technicalities, we assume that $N_{n,m}$ is finite in this work.
% We also would like to emphasize that the obtained results in this work holds for arbitrarily finite $N_{n,m}$, no matter how small or large $N_{n,m}$ is.

For the AoI at the device, if the device samples the physical process at slot $t$ (i.e., $s(t)=1$), then the AoI decreases to one (due to one slot used for status sampling), otherwise, the AoI increases by one. Then, the dynamics of the AoI at the device will be given by:
\begin{equation}\label{eqn:aoidevice}
A_l(t+1)=\begin{cases}1, & \text{if}~s(t)=1, \\
                \min\{A_l(t)+1,\hat{A}_l\},  &\text{otherwise.}
      \end{cases}.
\end{equation}

For the AoI at the destination, if the device sends the status packet to the destination at slot $t$ (i.e., $u(t)=1$), then the AoI decreases to the AoI at the device at slot $t$ plus one (due to one slot used for status packet transmission), otherwise, the AoI increases by one. Then, the dynamics of the AoI at the destination will be given by:
\begin{equation}\label{eqn:aoibs}
A_r(t+1)=\begin{cases}\min\{A_l(t)+1,\hat{A}_r\}, & \text{if}~u(t)=1, \\
                \min\{A_r(t)+1,\hat{A}_r\},  &\text{otherwise.}
      \end{cases}.
\end{equation}
\textcolor{black}{Note that, the analytical framework can be extended to the scenario in which  more than one slot are needed to generate or send  a status packet, by modifying the AoI dynamics in \eqref{eqn:aoidevice} and \eqref{eqn:aoibs}, accordingly.}
Clearly, it may not be optimal for the device to sample the physical process immediately after updating the status. The reason is that the newly generated status packet, if not transmitted to the destination  \textcolor{black}{immediately (due to a possibly poor channel state)}, can become stale and less useful for the destination, yielding energy waste \textcolor{black}{for sampling}.
Therefore, we are motivated to investigate how to jointly control the sampling and updating processes so as to minimize the AoI at the destination,  under the stringent energy constraint at the IoT device.

\subsection{CMDP Formulation and Optimality Equation}
\subsubsection{CMDP Formulation}
Given an observed system AoI state $\bs{A}$ and channel state $h$, the IoT device determines the sampling and updating action $\bs{w}$ according to the following policy.\footnote{\textcolor{black}{Here, we consider the entire AoI state space of $\bs{A}$. However, in practice, one may only consider the AoI states $\bs{A}$ such that $A_r\geq A_l$ without sacrificing optimality.}}
\begin{definition}\label{definition:stationary_policy}
A \emph{stationary sampling and updating policy} $\pi$ is defined as a mapping from the system AoI and the channel states $(\bs{A},h)\in\mathcal{A}\times\mathcal{H}$ to the control action of the device $\bs{w}\in\mathcal{W}$, where $\pi(\bs{A},h)=\bs{w}$.
\end{definition}

Under the i.i.d. assumption for the channel state process and the AoI dynamics in \eqref{eqn:aoidevice} and \eqref{eqn:aoibs}, the induced random process $\{(\bs{A}(t),h(t))\}$ is a controlled Markov chain. Hereinafter, as is commonly done  (e.g., see \cite{djonin2007mimo} and \cite{wcncage}), we restrict our attention to stationary unichain policies to guarantee the existence of the stationary optimal policies.
For a given stationary unichain policy $\pi$, the average AoI at the destination and the average energy cost will be:
% \begin{equation}
%    \bar{A}_r(\pi)\triangleq\limsup_{T\to\infty}\frac{1}{T}\sum_{t=1}^T \mathbb{E} \left[A_r(t)\right],\label{eqn:avg_aoi}
% \end{equation}
% \begin{equation}
%    \bar{C}(\pi)\triangleq\limsup_{T\to\infty}\frac{1}{T}\sum_{t=1}^T \mathbb{E} \left[C(\bs{w}(t))\right], \label{eqn:avg_cost}
% \end{equation}
\begin{align}
   &\bar{A}_r(\pi)\triangleq\limsup_{T\to\infty}\frac{1}{T}\sum_{t=1}^T \mathbb{E} \left[A_r(t)\right],\label{eqn:avg_aoi}\\
   &\bar{C}(\pi)\triangleq\limsup_{T\to\infty}\frac{1}{T}\sum_{t=1}^T \mathbb{E} \left[C(\bs{w}(t))\right], \label{eqn:avg_cost}
\end{align}
   % $$\bar{A}_r(\pi)\triangleq\limsup_{T\to\infty}\frac{1}{T}\sum_{t=1}^T \mathbb{E} \left[A_r(t)\right],$$
   % $$\bar{C}(\pi)\triangleq\limsup_{T\to\infty}\frac{1}{T}\sum_{t=1}^T \mathbb{E} \left[C(\bs{w}(t))\right], $$
where the expectation is taken with respect to the measure induced by the policy $\pi$.

We seek to find the optimal sampling and updating policy that minimizes the average AoI at the destination under an average energy cost constraint at the device, as follows:
% \begin{problem}\label{problem:cmdp}
% \begin{align}
% &\bar{A}_r^*\triangleq\min_{\pi}\bar{A}_r(\pi),\label{eqn:opt_Ar}\\
% &\text{s.t.~} \bar{C}(\pi) \leq C^{\textrm{max}}.\label{eqn:constraint}
% \end{align}
% where $\pi$ is a stationary unchain policy in Definition~\ref{definition:stationary_policy} and $\bar{A}_r^*$ denotes the minimum average AoI at the destination achieved by the optimal policy $\pi^*$ under the constraint in \eqref{eqn:constraint}.
% \end{problem}
\begin{subequations}\label{eqn:cmdp}
\begin{align}
&\bar{A}_r^*\triangleq\min_{\pi}\bar{A}_r(\pi),\label{eqn:opt_Ar}\\
&\text{s.t.~} \bar{C}(\pi) \leq C^{\textrm{max}}.\label{eqn:constraint}
\end{align}
\end{subequations}
Here $\pi$ is a stationary unichain policy and $\bar{A}_r^*$ denotes the minimum average AoI at the destination achieved by the optimal policy $\pi^*$ under the constraint in \eqref{eqn:constraint}.
The problem in \eqref{eqn:cmdp} is an infinite horizon average cost CMDP, which is to known to be challenging due to the curse of dimensionality.
% One can easily see that, for any $h\in\mathcal{H}$ and any $\bs{A}\in\mathcal{A}$ satisfying $A_l=A_r$, it is not optimal for the IoT device to send the status update, i.e., $\pi_u^*(\bs{A},h)=0$.
% In the following proposition, we show that it is not optimal for the IoT device to send the status update when the AoI at the destination equals to the AoI at the device.
% The reason is that, the status packet currently available at the device has already been sent to the destination previously.
% % This can be seen from the AoI dynamics in \eqref{eqn:aoidevice} and \eqref{eqn:aoibs}.
%  The proof of the proposition is straightforward and, thus, omitted.
% \begin{proposition}\label{prop:prop}
% For any $h\in\mathcal{H}$ and any $\bs{A}\in\mathcal{A}$ satisfying $A_l=A_r$, the optimal updating policy is $\pi_u^*(\bs{A},h)=0$.
% \end{proposition}
\subsubsection{Optimality Equation}
To obtain the optimal policy $\pi^*$ for the CMDP in \eqref{eqn:cmdp}, we reformulate the CMDP into a parameterized unconstrained MDP using the Lagrangian approach\cite{cmdp}. For a given Lagrange multiplier $\lambda$, we define the Lagrange cost at slot $t$ as
\begin{equation}
L(\bs{A}(t),h(t),\bs{w}(t);\lambda)\triangleq A_r(t) + \lambda C(\bs{w}(t)).
\end{equation}
Then, the average Lagrange cost under policy $\pi$ is given by:
\begin{align}
   \bar{L}(\pi;\lambda)\triangleq\limsup_{T\to\infty}\frac{1}{T}\sum_{t=1}^T \mathbb{E} \left[L(\bs{A}(t),h(t),\bs{w}(t);\lambda)\right].\label{eqn:avg_mdp}
\end{align}
Now, we have an unconstrained MDP whose goal is to minimize the average Lagrange cost:
% \begin{problem}\label{problem:mdp}
% \begin{align}
% \bar{L}^*(\lambda)\triangleq\min_{\pi}\bar{L}(\pi;\lambda),\label{eqn:opt_L}
% \end{align}
% where  $\bar{L}^*(\lambda)$ denotes the minimum average Lagrange cost achieved by the optimal policy $\pi^*_{\lambda}$ for a given $\lambda$.
% \end{problem}
% \begin{subequations}\label{eqn:mdp}
\begin{align}
\bar{L}^*(\lambda)\triangleq\min_{\pi}\bar{L}(\pi;\lambda),\label{eqn:opt_L}
\end{align}
% \end{subequations}
where  $\bar{L}^*(\lambda)$ is the minimum average Lagrange cost achieved by the optimal policy $\pi^*_{\lambda}$ for a given $\lambda$.
% According to \cite[Theorem 12.7]{cmdp} and by using the results in \cite{BEUTLER1985236},
According to\textcolor{black}{\cite[Theorem 1]{djonin2007mimo} and \cite[Theorem 4.4]{BEUTLER1985236},}
 we have the following relation between the optimal solutions of the problems in \eqref{eqn:cmdp} and \eqref{eqn:opt_L}.
\begin{lemma} \label{lemma:relation} The optimal average AoI cost in \eqref{eqn:opt_Ar} and the optimal average Lagrange cost in \eqref{eqn:opt_L} satisfy:
\begin{equation}
\bar{A}_r^* = \max_{\lambda\geq 0} \bar{L}^*(\lambda) - \lambda C^{\textrm{max}}.
\end{equation}
The optimal policy $\pi^*$ of the CMDP in \eqref{eqn:cmdp}  is a randomized mixture of two deterministic stationary policies $\pi^*_{\lambda_1}$ and $\pi^*_{\lambda_2}$, in the form of
\begin{equation}\label{eqn:optimal_pi}
\pi^* = \alpha \pi^*_{\lambda_1} + (1-\alpha) \pi^*_{\lambda_2},
\end{equation}
where $\alpha\in[0,1]$ is the randomization parameter, and $\pi^*_{\lambda_1}$ and $\pi^*_{\lambda_2}$ are the optimal policies of the unconstrained MDP in \eqref{eqn:opt_L}
under the Lagrange multipliers $\lambda_1$ and $\lambda_2$, respectively.
\end{lemma}

To obtain the optimal policy $\pi^*$ of the CMDP, according to \cite[Propositions 4.2.1, 4.2.3, and 4.2.5]{bertsekas} \textcolor{black}{(these propositions are restated in Appendix H)}, we first obtain the optimal policy $\pi^*_{\lambda}$ for a given $\lambda$ of the unconstrained MDP by solving the following Bellman equation.
\begin{lemma}\label{lemma:bellman}
For any $\lambda$, there exists $(\theta_{\lambda},\{V(\bs{A},h;\lambda)\})$ satisfying:
% \begin{align}
%   \theta_{\lambda}&+V(\bs{A},h;\lambda)=\min_{\bs{w}\in\mathcal{W}}\Big\{L(\bs{A},h,\bs{w};\lambda)\nonumber\\&\hspace{1mm}+\sum_{h'\in\mathcal{H}}p_{\mathcal{H}}(h')V(\bs{A}',h';\lambda)\Big\},
%   ~\forall (\bs{A},h)\in\mathcal{A}\times\mathcal{H},\label{eqn:bellman}
% \end{align}
\begin{align}
  \theta_{\lambda}+V(\bs{A},h;\lambda)=\min_{\bs{w}\in\mathcal{W}}\left\{L(\bs{A},h,\bs{w};\lambda)+\sum_{h'\in\mathcal{H}}p_{\mathcal{H}}(h')V(\bs{A}',h';\lambda)\right\},
  ~\forall (\bs{A},h)\in\mathcal{A}\times\mathcal{H},\label{eqn:bellman}
\end{align}
where $\bs{A}'$ satisfies the AoI dynamics in \eqref{eqn:aoidevice} and \eqref{eqn:aoibs},
$\theta_{\lambda}=\bar{L}^*(\lambda)$ is the optimal value to \eqref{eqn:opt_L} for all initial state $(\bs{A}(1),h(1))$, and $V(\cdot)$ is the value function \textcolor{black}{which is a mapping from $(\bs{A},h)$ to real values.}
Moreover, for a given $\lambda$, the optimal policy achieving  $\bar{L}^*(\lambda)$ will be
% \begin{align}
%   \pi^*_{\lambda}&(\bs{A},h)=\arg\min_{\bs{w}\in\mathcal{W}}\Big\{L(\bs{A},h,\bs{w};\lambda)\nonumber\\&\hspace{1mm}+\sum_{h'\in\mathcal{H}}p_{\mathcal{H}}(h')V(\bs{A}',h';\lambda)\Big\},\forall (\bs{A},h)\in\mathcal{A}\times\mathcal{H}.\label{eqn:pi_lambda}
% \end{align}
\begin{align}
  \pi^*_{\lambda}(\bs{A},h)=\arg\min_{\bs{w}\in\mathcal{W}}\left\{L(\bs{A},h,\bs{w};\lambda)+\sum_{h'\in\mathcal{H}}p_{\mathcal{H}}(h')V(\bs{A}',h';\lambda)\right\},\forall (\bs{A},h)\in\mathcal{A}\times\mathcal{H}.\label{eqn:pi_lambda}
\end{align}
\end{lemma}

From Lemma~\ref{lemma:bellman}, we can see that $\pi^*_{\lambda}$ , which is given by \eqref{eqn:pi_lambda}, depends on $(\bs{A},h)$ through the value function $V(\cdot)$. Determining $V(\cdot)$ involves solving the Bellman equation in \eqref{eqn:bellman}, for which there is no closed-form solution in general \cite{bertsekas}. Numerical algorithms such as value iteration and policy iteration are usually \textcolor{black}{computationally} impractical to implement for an IoT due to the curse of dimensionality and they do not typically yield many design insights. Therefore, it is desirable to analyze the structural properties of $\pi^*_{\lambda}$, as we do next.

\subsection{Structural Analysis and Algorithm Design}
First, we characterize the structural properties of $\pi^*_{\lambda}$ for the unconstrained MDP in \eqref{eqn:opt_L}. Then, we propose a novel structure-aware optimal algorithm to obtain the optimal policy $\pi^*$ for the CMDP in \eqref{eqn:cmdp}.
Finally, we study the effects of the wireless channel fading.
\subsubsection{Optimality Properties}

By using the relative value iteration algorithm and the special structures of the AoI dynamics in \eqref{eqn:aoidevice} and \eqref{eqn:aoibs}, we can prove the following property.

\begin{lemma}\label{lemma:valuefunction}
Given $\lambda\geq 0$, $V(\bs{A},h;\lambda)$ is non-decreasing with $A_l$ and $A_r$ for any $h\in\mathcal{H}$.
\end{lemma}
\begin{IEEEproof}
  % Please see Appendix~\ref{app:value_function}.
  See Appendix~A.
\end{IEEEproof}

Then, we introduce the state-action Lagrange cost function, which is related to the right-hand side of the Bellman equation in \eqref{eqn:bellman} and is given by:
\begin{equation}
J(\bs{A},h,\bs{w};\lambda)\triangleq L(\bs{A},h,\bs{w};\lambda)+\sum_{h'\in\mathcal{H}}p_{\mathcal{H}}(h')V(\bs{A}',h';\lambda).\label{eqn:state-action-function}
\end{equation}
We now define $\Delta J_{\bs{w},\bs{w}'}(\bs{A},h;\lambda)\triangleq J(\bs{A},h,\bs{w};\lambda)-J(\bs{A},h,\bs{w}';\lambda)$.
% \begin{equation}
% \Delta J_{\bs{w},\bs{w}'}(\bs{A},h;\lambda)\triangleq J(\bs{A},h,\bs{w};\lambda)-J(\bs{A},h,\bs{w}';\lambda).
% \end{equation}
If $\Delta J_{\bs{w},\bs{w}'}(\bs{A},h;\lambda)\leq 0$, we say that action $\bs{w}$ \emph{dominates} action $\bs{w}'$ at state $(\bs{A},h)$ for a given $\lambda$.
By Lemma~\ref{lemma:valuefunction}, if $\bs{w}$ dominates all other actions at state $(\bs{A},h)$ for a given $\lambda$, then we have $\pi^*_{\lambda}(\bs{A},h)=\bs{w}$.
Based on Lemma~\ref{lemma:valuefunction}, we can obtain the following properties of $\Delta J_{\bs{w},\bs{w}'}(\bs{A},h;\lambda)$.

\begin{lemma}\label{lemma:stateactionfunction}
Given $\lambda\geq 0$, for any $\bs{A}\in\mathcal{A}$, $h\in\mathcal{H}$, and $\bs{w},\bs{w}'\in\mathcal{W}$, $\Delta J_{\bs{w},\bs{w}'}(\bs{A},h;\lambda)$ has the following properties:
\begin{enumerate}[label=\Alph*)]
  \item If $\bs{w}=(0,0)$, then $\Delta J_{\bs{w},\bs{w}'}(\bs{A},h;\lambda)$ is non-decreasing with $A_l$ for $\bs{w}'=(1,0)$ and non-decreasing with $A_r$ for \textcolor{black}{$\bs{w}'=(0,1) \text{~or~} (1,1)$}.
  \item If $\bs{w}=(0,1)\text{~or~}(1,1)$, then  $\Delta J_{\bs{w},\bs{w}'}(\bs{A},h;\lambda)$ is non-increasing with $A_r$ for any $\bs{w}'\neq \bs{w}$.   % $\bs{w}'=(0,0), (1,0)$.
  \item If $\bs{w}=(1,0)$, then  $\Delta J_{\bs{w},\bs{w}'}(\bs{A},h;\lambda)$ is non-increasing with $A_l$ for any $\bs{w}'\neq \bs{w}$. %$\bs{w}'=(0,1), (1,1)$.
\end{enumerate}
\end{lemma}
\begin{IEEEproof}
  % Please see Appendix~\ref{app:state_action_function}.
  See Appendix~B.
\end{IEEEproof}

Lemma~\ref{lemma:stateactionfunction} follows from the special properties of the AoI dynamics and is essential for the characterization of the structural properties of $\pi^*_{\lambda}$.
The property shown in Lemma~\ref{lemma:stateactionfunction} is similar to the diminishing-return property of multimodularity functions \cite{Koole}.
% From Lemma~\ref{lemma:stateactionfunction}, we can see that for a given $\lambda$ and $h$, if action $(0,0)$ dominates action $(0,1)$ or $(1,1)$ for some AoI $(A_l,A_r)$, then  $(0,0)$ still dominates $(0,1)$ or $(1,1)$ for state $(A_l,A_r-1)$; if action $(0,0)$ dominates action $(1,0)$ for some AoI $(A_l,A_r)$, then  $(0,0)$ still dominates $(1,0)$ for $(A_l-1,A_r)$; if $(1,0)$
 From Lemma~\ref{lemma:stateactionfunction}, we can see that for a given $\lambda$ and $h$, if action $\bs{w}$ dominates action $\bs{w}'$ for some AoI state $\bs{A}$, then  $\bs{w}$ still dominates $\bs{w}'$ for another AoI $\bs{A}'$, provided that  $\bs{A}$ and $\bs{A}'$ satisfy certain conditions \textcolor{black}{such that $\Delta J_{\bs{w},\bs{w}'}(\bs{A}',h;\lambda)\leq \Delta J_{\bs{w},\bs{w}'}(\bs{A},h;\lambda)\leq 0$.} Before presenting the structure of  $\pi^*_{\lambda}$ in Theorem~\ref{theorem:optimal}, we make the following definitions:
% \begin{align}
% &\Phi_{\bs{w}}(A_r,h;\lambda)\triangleq \{A_l|A_l\in\mathcal{A}_l\text{~and~}\Delta J_{\bs{w},\bs{w}'}(\bs{A},h;\lambda)\leq 0\nonumber\\&~\hspace{30mm}~\forall \bs{w}'\in\mathcal{W}\text{~and~}\bs{w}'\neq \bs{w}\},\\
% &\Psi_{\bs{w}}(A_l,h;\lambda)\triangleq \{A_r|A_r\in\mathcal{A}_r\text{~and~}\Delta J_{\bs{w},\bs{w}'}(\bs{A},h;\lambda)\leq 0\nonumber\\&~\hspace{30mm}~\forall \bs{w}'\in\mathcal{W}\text{~and~}\bs{w}'\neq \bs{w}\}.
% \end{align}
\begin{align}
&\Phi_{\bs{w}}(A_r,h;\lambda)\triangleq \left\{A_l|A_l\in\mathcal{A}_l\text{~and~}\Delta J_{\bs{w},\bs{w}'}(\bs{A},h;\lambda)\leq 0~\forall \bs{w}'\in\mathcal{W}\text{~and~}\bs{w}'\neq \bs{w}\right\},\\
&\Psi_{\bs{w}}(A_l,h;\lambda)\triangleq \left\{A_r|A_r\in\mathcal{A}_r\text{~and~}\Delta J_{\bs{w},\bs{w}'}(\bs{A},h;\lambda)\leq 0~\forall \bs{w}'\in\mathcal{W}\text{~and~}\bs{w}'\neq \bs{w}\right\}.
\end{align}
Then, we define:
\begin{align}
&\phi_{\bs{w}}^+(A_r,h;\lambda)\triangleq\begin{cases}\max\Phi_{\bs{w}}(A_r,h;\lambda),  & \text{if}~\Phi_{\bs{w}}(A_r,h;\lambda)\neq\emptyset, \\
            -\infty,  &\text{otherwise},
  \end{cases}\label{eqn:func1}\\
&\phi_{\bs{w}}^-(A_r,h;\lambda)\triangleq\begin{cases}\min\Phi_{\bs{w}}(A_r,h;\lambda),  & \text{if}~\Phi_{\bs{w}}(A_r,h;\lambda)\neq\emptyset, \\
            +\infty,  &\text{otherwise},
  \end{cases}\label{eqn:func2}\\
&\psi_{\bs{w}}^+(A_l,h;\lambda)\triangleq\begin{cases}\max\Psi_{\bs{w}}(A_l,h;\lambda),  & \text{if}~\Psi_{\bs{w}}(A_l,h;\lambda)\neq\emptyset, \\
            -\infty,  &\text{otherwise},
  \end{cases}\label{eqn:func3}\\
&\psi_{\bs{w}}^-(A_l,h;\lambda)\triangleq\begin{cases}\min\Psi_{\bs{w}}(A_l,h;\lambda),  & \text{if}~\Psi_{\bs{w}}(A_l,h;\lambda)\neq\emptyset, \\
            +\infty,  &\text{otherwise}.
  \end{cases}\label{eqn:func4}
\end{align}
% Now, we can show the following theorem.
\begin{theorem} \label{theorem:optimal}
Given $\lambda$, for any $\bs{A}\in\mathcal{A}$ and $h\in\mathcal{H}$, \textcolor{black}{there exists an optimal policy satisfying the following structural properties:}
\begin{enumerate}[label=\Alph*)]
  \item $\pi^*_{\lambda}(\bs{A},h)=(0,0)$, for all $\bs{A}\in\mathcal{A}_0(h;\lambda)\triangleq\{\bs{A}|A_l\leq \phi_{(0,0)}^+(A_r,h;\lambda), A_r\leq \psi_{(0,0)}^+(A_l,h;\lambda)\}$.
  \item $\pi^*_{\lambda}(\bs{A},h)=(0,1)$ if $A_r\geq \psi_{(0,1)}^-(A_l,h;\lambda)$.
  \item $\pi^*_{\lambda}(\bs{A},h)=(1,0)$ if $A_l\geq \phi_{(1,0)}^-(A_r,h;\lambda)$.
  \item $\pi^*_{\lambda}(\bs{A},h)=(1,1)$ if $A_r\geq \psi_{(1,1)}^-(A_l,h;\lambda)$.
\end{enumerate}
\end{theorem}

\begin{figure}[!t]
\begin{centering}
\includegraphics[scale=.6]{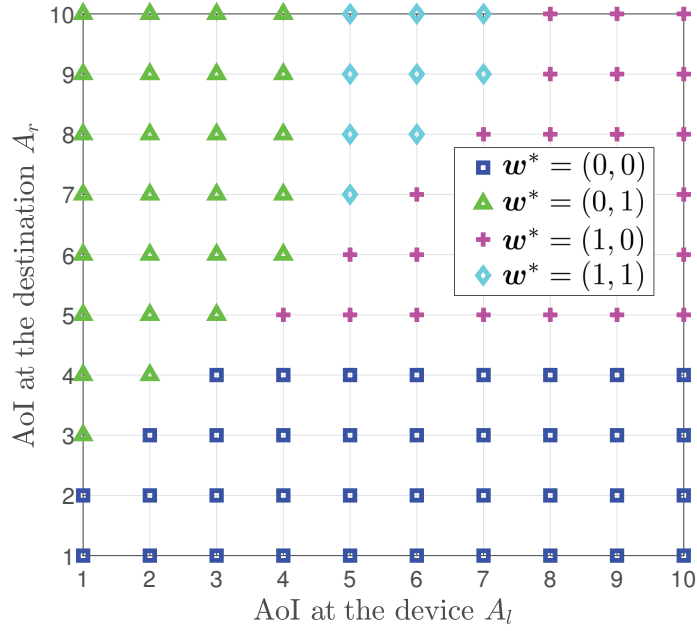}
 \caption{Structure of the optimal policy $\pi^*_{\lambda}$ for a given Lagrange multiplier $\lambda$ and channel state $h$. $\hat{A}_l=\hat{A}_r=10$. \textcolor{black}{$C_s=2$, $C_u(h)=3.5/h$, where $h=1,2$, \textcolor{black}{$C^{\max}=3$}}.}\label{fig:structure}
\end{centering}
% \vspace{-1cm}
\end{figure}

Theorem~\ref{theorem:optimal} characterizes the structural properties of the optimal policy $\pi^*_{\lambda}$ of the unconstrained MDP in \eqref{eqn:opt_L} for a given $\lambda$.
Fig.~\ref{fig:structure} illustrates the analytical results of Theorem~\ref{theorem:optimal}, where the optimal policy is computed numerically using policy iteration \cite[Chapter 8.6]{puterman}.
Fig.~\ref{fig:structure} shows that, if the AoI state falls into the region of the black squares (i.e., $\mathcal{A}_0(h;\lambda)$), the device will remain idle and will not sample the physical process nor send the status update. Thus, $\mathcal{A}_0(h;\lambda)$ is referred to as the \emph{idle region}.
For given $A_l$, $h$, and $\lambda$, \textcolor{black}{the scheduling of} $(0,1)$ (or $(1,1)$) is threshold-based with respect to $A_r$.\
In other words, when $A_r$ is small, it is not efficient to send a new status update to the destination, as a higher updating cost per age is consumed. Meanwhile, when $A_r$ is large, it is more efficient to update the status, as the status packet at the destination becomes more outdated.
For given $A_r$, $h$, and $\lambda$, \textcolor{black}{the scheduling of} $(1,0)$ is threshold-based with respect to $A_l$. \
Hence when $A_l$ is small, it is not efficient to sample the physical process, as a higher sampling cost per age is incurred. In contrast, when $A_l$ is large, it is more efficient to generate a new status packet, as the status packet at the device becomes more outdated and less useful for the destination.
\textcolor{black}{These observations indicate that the zero-wait policy (i.e., transmit the status packet immediately after sampling) may be detrimental to the minimization of the AoI, because of the energy cost constraint.
This is reminiscent of the result in [11], however, the work in [11] obtained this outcome because of the considered random service times.}
% \textcolor{black}{These observations are also consistent with the results that the zero-wait policy (i.e., transmit the status packet immediately after sampling) may be detrimental to the minimization of the AoI \cite{8000687}.}
\emph{These threshold-based properties reveal a fundamental tradeoff between the AoI \textcolor{black}{at} the destination and the sampling and updating costs.}
Such structural properties provide valuable insights for the design of the sampling and updating processes in practical IoT systems.
Here, we would like to emphasize that, although the threshold-based structures may look intuitive, it is challenging to prove these structures rigorously. This is due to the coupled two AoI states and the special AoI dynamics.
\textcolor{black}{Moreover, it is not always possible to fully characterize the structural properties of the optimal policy, e.g., the structure with respect to $h$ and the structures of the thresholds, as the (generally required) key property of the value function, i.e., the multimodularity\cite{Koole}, does not hold for our value function.}

\subsubsection{Algorithm Design}
By exploiting the results of Theorem~\ref{theorem:optimal}, we first propose a structure-aware algorithm to compute the optimal policy $\pi^*_{\lambda}$ for a given $\lambda$, and, then, we describe how to update $\lambda$ and obtain the optimal policy $\pi^*$.
\textcolor{black}{Note that, although the exact values of the thresholds in Theorem~\ref{theorem:optimal} rely on the exact values of $V(\bs{A},h;\lambda)$, the threshold-based structure only relies on the properties of $V(\bs{A},h;\lambda)$ and $J(\bs{A},h,\bs{w};\lambda)$. These properties can be exploited to reduce the computational complexity for obtaining the optimal policy, without knowing the exact values of the thresholds. In particular,}
by properties B)-D) in Theorem~\ref{theorem:optimal}, we know that the optimal action for a certain system state is still optimal for some other system states. In particular, we can see that, for all $\bs{A}$ and $h$,
\begin{equation}\label{eqn:imply}
\begin{cases}
&\pi^*_{\lambda}(A_l,A_r,h)=(0,1)~\Rightarrow~\pi^*_{\lambda}(A_l,A_r+1,h)=(0,1),\\
&\pi^*_{\lambda}(A_l,A_r,h)=(1,0)~\Rightarrow~\pi^*_{\lambda}(A_l+1,A_r,h)=(1,0),\\
&\pi^*_{\lambda}(A_l,A_r,h)=(1,1)~\Rightarrow~\pi^*_{\lambda}(A_l,A_r+1,h)=(1,1).
\end{cases}
\end{equation}
Therefore, to find $\pi^*_{\lambda}$, we only need to minimize in the right-hand side of \eqref{eqn:pi_lambda} for some $\bs{A}$, rather than for all $\bs{A}$, which reduces the computational complexity.
By incorporating  \eqref{eqn:imply} into a standard policy iteration algorithm, we can develop a structure-aware policy iteration algorithm, \textcolor{black}{as shown in Algorithm~\ref{alg:SPIA}}.
\textcolor{black}{It can be seen that Algorithm~\ref{alg:SPIA} is a monotone policy iteration algorithm (see an example in \cite[Chapter 8.11.2]{puterman}), and thus,  converges to the optimal policy $\pi^*_{\lambda}$\cite[Theorem 8.6.6]{puterman} \textcolor{black}{(restated in  Appendix H}).}
% According to \cite[Theorem 8.6.6]{puterman} \textcolor{black}{(given in Appendix H)}, we know that our proposed structure-aware policy iteration algorithm converges to the optimal policy $\pi^*_{\lambda}$ and, thus, is an optimal algorithm.
Note that, when one of the ``if'' conditions in \textcolor{black}{Step~\ref{eqn:spia_imp} of Algorithm~\ref{alg:SPIA}} is satisfied for  a certain system state, we can determine the optimal action immediately, \textcolor{black}{without performing the minimization in \eqref{eqn:update_pi}.
The computational complexity saving for each iteration in Algorithm~\ref{alg:SPIA} is $O(|\mathcal{M}|(|\mathcal{A}||\mathcal{H}|)^2)$\cite{complexity}. This is reasonable since the complexity saving grows exponentially with the state space.} 

\begin{algorithm}[!t]
% \small
\caption{\textcolor{black}{Structure-aware Policy Iteration Algorithm}}
\label{alg:SPIA}
\begin{algorithmic}[1]
\State Set $\pi^*_{\lambda,0}(\bs{A},h)=(0,0)$ for all $(\bs{A},h)\in\mathcal{A}\times\mathcal{H}$, select reference state $(\bs{A}^\dag,h^\dag)$, and set $m=0$.
\State (Policy Evaluation) Given policy $\pi^*_{\lambda,m}$, compute the value $\theta_{\lambda,m}$ and value function $V_{m}(\bs{A},h)$ from the linear system of equations\footnotemark
\begin{equation}\label{eqn:spia_eva}
\begin{cases}
   \theta_{\lambda,m}+V_m(\bs{A},h;\lambda)=L(\bs{A},h,\pi^*_{\lambda,m}(\bs{A},h);\lambda)+\sum_{h'\in\mathcal{H}}p_{\mathcal{H}}(h')V_m(\bs{A}',h';\lambda),\forall (\bs{A},h)\\
   V_m(\bs{A}^\dag,h^\dag;\lambda)=0
\end{cases},
\end{equation}
\Statex where $\bs{A}'$ satisfies the AoI dynamics in \eqref{eqn:aoidevice} and \eqref{eqn:aoibs} under the action $\pi^*_{\lambda,m}(\bs{A},h)$.\label{code:spia_eva}
\State (Structured Policy Improvement) Find a new policy $\pi^*_{\lambda,m+1}$ for each $\bs{A}\in\mathcal{A}$ and $h\in\mathcal{H}$, where for each $(\bs{A},h)\in\mathcal{A}\times\mathcal{H}$,  $\pi^*_{\lambda,m+1}(\bs{A},h;\lambda)$ is such that:\label{eqn:spia_imp}
\Statex\textbf{if} $\textcolor{black}{\pi^*_{\lambda,m+1}((A_l,A_r-1),h;\lambda)}=(0,1)$, \textbf{then} $\pi^*_{\lambda,m+1}(\bs{A},h;\lambda)=(0,1).$
        % \begin{equation*}
        %   \pi^*_{\lambda,m}(\bs{A},h;\lambda)=(0,1).
        % \end{equation*}
\Statex\textbf{else if} $\pi^*_{\lambda,m+1}((A_l-1,A_r),h;\lambda)=(1,0)$, \textbf{then} $\pi^*_{\lambda,m+1}(\bs{A},h;\lambda)=(1,0).$
        % \begin{equation*}
        %   \pi^*_{\lambda,m}(\bs{A},h;\lambda)=(1,0).
        % \end{equation*}
\Statex\textbf{else if} $\pi^*_{\lambda,m+1}((A_l,A_r-1),h;\lambda)=(1,1)$, \textbf{then} $\pi^*_{\lambda,m+1}(\bs{A},h;\lambda)=(1,1).$
        % \begin{equation*}
        %   \pi^*_{\lambda,m}(\bs{A},h;\lambda)=(1,1).
        % \end{equation*}
\Statex\textbf{else}
      % \begin{align}\label{eqn:update_pi}
      % \pi^*_{\lambda,m}(\bs{A},h;\lambda)&=\arg\min_{\bs{w}\in\mathcal{W}}\bigg\{L(\bs{A},h,\bs{w};\lambda)\nonumber\\&\hspace{1mm}+\sum_{h'\in\mathcal{H}}p_{\mathcal{H}}(h')V_{m-1}(\bs{A}',h';\lambda)\bigg\}.
      % \end{align}
      \begin{align}\label{eqn:update_pi}
      \hspace{-1cm}\pi^*_{\lambda,m+1}(\bs{A},h;\lambda)=\arg\min_{\bs{w}\in\mathcal{W}}\bigg\{L(\bs{A},h,\bs{w};\lambda)+\sum_{h'\in\mathcal{H}}p_{\mathcal{H}}(h')V_{m}(\bs{A}',h';\lambda)\bigg\}.
      \end{align}
\State Go to Step \ref{code:spia_eva} until $\mu_{l+1}^*=\mu_{l}^*$.
\end{algorithmic}
\end{algorithm}

\footnotetext{\textcolor{black}{The solution to \eqref{eqn:spia_eva} can be derived  using Gaussian elimination or  the relative value iteration method\cite{bertsekas}.}}

From \eqref{eqn:optimal_pi}, we know that obtaining $\pi^*$ requires computing the two Lagrange multipliers $\lambda_1$ and $\lambda_2$, and the randomization parameter $\alpha$.
As in \cite{BEUTLER1985236} and \cite{5165031}, we set $\lambda_1=\lambda^*-\eta$ and $\lambda_2=\lambda^*+\eta$, where the perturbation parameter $\eta$ is some small constant and $\lambda^*$ is the optimal Lagrange multiplier satisfying $\lambda^*=\min\{\lambda:\bar{C}(\pi^*_{\lambda}) \leq C^{\textrm{max}}\}$. \textcolor{black}{By using the Robbins-Monro algorithm\cite{robbins1951}, which is a stochastic gradient-based algorithm}, the Lagrange multiplier is updated according to
% \begin{equation}
$\lambda_{m+1}= \lambda_m + \epsilon_m\left(\bar{C}(\pi^*_{\lambda_m})-C^{\textrm{max}}\right)$,
% \end{equation}
where the step $\epsilon_m=\frac{1}{m}$ and $\lambda_{1}$ is initialized with a sufficiently large number.
The generated sequence $\{\lambda_m\}$ converges to the optimal Lagrange multiplier $\lambda^*$\cite{robbins1951}.
Then, the randomization parameter $\alpha$ is given by: $\alpha=(C^{\textrm{max}}-\bar{C}(\pi^*_{\lambda_2}))\slash (\bar{C}(\pi^*_{\lambda_1})-\bar{C}(\pi^*_{\lambda_2}))$.
% \begin{equation}
% \alpha=\frac{C^{\textrm{max}}-\bar{C}(\pi^*_{\lambda_2})}{\bar{C}(\pi^*_{\lambda_1})-\bar{C}(\pi^*_{\lambda_2})}.
% \end{equation}
\textcolor{black}{Here, $\alpha$ is chosen such that $\alpha \bar{C}(\pi^*_{\lambda_1}) + (1-\alpha)\bar{C}(\pi^*_{\lambda_2})= C^{\textrm{max}}$. Then, for some perturbation parameter $\eta$, by \cite[Theorem 4.3]{BEUTLER1985236}, we have that $
\pi^* = \alpha \pi^*_{\lambda_1} + (1-\alpha) \pi^*_{\lambda_2}$ is the optimal policy for the CMDP.
Note that, $\alpha$ is guaranteed to lie in $(0,1)$ since $\bar{C}(\pi^*_{\lambda})$ is non-increasing with $\lambda$\cite{BEUTLER1985236}.}
% In Fig. , we outline the overall algorithm for solving Problem~\ref{problem:cmdp}.
So far, we have characterized the structural properties of the optimal policy $\pi^*_{\lambda}$ and developed a structure-aware optimal algorithm.

\subsubsection{Effects of Wireless Channel Dynamics}
Next, we study the influence of the wireless channel fading distribution on the optimal average AoI at the destination.  The results are established by using the stochastic dominance relations of random variables. From \cite{djonin2007mimo} and \cite{OR}, we present the following definition.

\begin{definition}\label{definition:stochastis_dominance}
Let $x(\gamma)$ be a random variable with the support on the set $\mathcal{X}$ according to a probability measure $\mu(\gamma)$ parameterized by some $\gamma$. $x(\gamma_1)$ is said to \emph{stochastically dominate} $x(\gamma_2)$ on the set of functions $\mathcal{F}$, or $x(\gamma_1)\succeq_{\mathcal{F}}x(\gamma_2)$, if $\mathbb{E}[f(x(\gamma_1))]\geq \mathbb{E}[f(x(\gamma_2))]$,
% \begin{equation}
% \mathbb{E}[f(x(\gamma_1))]\geq \mathbb{E}[f(x(\gamma_2))],
% \end{equation}
for all functions $f\in\mathcal{F}$.
If $\mathcal{F}$ is the set of increasing functions, then $\succeq_{\mathcal{F}}$ corresponds to the \emph{first-order stochastic dominance}.
If $\mathcal{F}$ is the set of increasing and concave functions, then $\succeq_{\mathcal{F}}$ corresponds to the \emph{second-order stochastic dominance}.
\end{definition}

Consider two channels $I$ and $J$. Let $h^I\in\mathcal{H}$  and $h^J\in\mathcal{H}$ be random variables with the fading distributions $p_{\mathcal{H}}^I(h)$ and $p_{\mathcal{H}}^J(h)$ for channels $I$ and $J$, respectively.
% Consider two channels $h^I\in\mathcal{H}$ and $h^J\in\mathcal{H}$, with the associated channel fading distributions $p_{\mathcal{H}}^I(h)$ and $p_{\mathcal{H}}^J(h)$.
% In this section, we explicitly state the dependence of AoI, costs, and value functions by using the appropriate superscripts $I$ and $J$, wherever needed.
\begin{theorem}\label{theorem:first_order}
If $h^I$ first-order stochastically dominates $h^J$, then we have
\begin{equation}
\bar{A}_r^{I*} \leq \bar{A}_r^{J*},
\end{equation}
where $\bar{A}_r^{I*}$ and $\bar{A}_r^{J*}$ are the optimal average AoI at the destination under channels $I$ and $J$, respectively.
\end{theorem}

\begin{IEEEproof}
See Appendix D.
\end{IEEEproof}

Theorem~\ref{theorem:first_order} demonstrates that channels with larger mean channel gain can achieve smaller AoI at the destination under the same resource constraint.
Following the proof of Theorem~\ref{theorem:first_order}, we have the following corollary for second-order stochastically dominating channels.
\begin{corollary}\label{corollary:second_order}
If $h^I$ second-order stochastically dominates $h^J$ and $C_u(h)$ is decreasing and convex with $h$, \textcolor{black}{then the optimal average AoI at the destination under channel $I$ is smaller than that under channel $J$, i.e., $\bar{A}_r^{I*} \leq \bar{A}_r^{J*}$.}
% \begin{equation}
% \bar{A}_r^{I*} \leq \bar{A}_r^{J*}.
% \end{equation}
\end{corollary}

Note that, for the transmission cost $C_u(h)$ defined according to the Shannon's formula (e.g., in \cite{djonin2007mimo}), it can be easily seen that $C_u(h)$ satisfies the conditions of Corollary~\ref{corollary:second_order}.
If  $h^I$ has the same mean as $h^J$, \textcolor{black}{by Definition~\ref{definition:stochastis_dominance},} the second-order stochastic dominance of $h^J$ over $C_u(h)$ indicates that  $h^I$ has smaller variance (i.e., less scattering) than $h^J$.
Therefore, Corollary~\ref{corollary:second_order} reveals that channels with less scattering and the same mean channel gain can achieve a smaller AoI at the destination under the same resource constraint.
\textcolor{black}{The results obtained in Theorem~\ref{theorem:first_order} and Corollary~\ref{corollary:second_order} reveal the fundamental monotone dependency of the optimal AoI at the destination on the transmission probability distribution of the CMDP in \eqref{eqn:cmdp}.}
% The results obtained in Theorem~\ref{theorem:first_order} and Corollary~\ref{corollary:second_order} can provide design insights from the perspective of cybersecurity.

Thus far, we have analyzed the optimality properties for the case of a single IoT device so that to gain a deep understanding of the behavior of the optimal sampling and updating policy for the real-time monitoring system. Next, we consider a more general scenario in which there are multiple IoT devices.
For such a scenario, the system state space is much larger than that for the case of a single IoT device, as it  grows exponentially with the number of the devices.
This hinders the structural analysis of the optimal policy and the design of an optimal algorithm with low-complexity.
Therefore, we will focus on the design of a low-complexity suboptimal solution for the case of multiple IoT devices.

\section{Semi-Distributed Suboptimal Sampling and Updating Control for Multiple IoT Devices}

\subsection{System Model and Problem Formulation}
We now extend the real-time monitoring system in Section II to a more general scenario, in which  a set $\mathcal{K}$ of $K$ IoT devices sample the associated physical processes and update the status packets to a common destination.
Hereinafter, with some notation abuse, for each IoT device $k\in\mathcal{K}$, we denote by $\bs{A}_k(t)\triangleq(A_{l,k}(t),A_{r,k}(t))\in\mathcal{A}_k$, $h_{k}(t)$, and $\bs{w}_k(t)\triangleq(s_k(t),u_k(t))\in\mathcal{W}_k$  the AoI state, the channel state,  and  the control action vector at slot $t$, respectively.
Under action $\bs{w}_k(t)$, the AoI state $\bs{A}_k(t)$ for each IoT device $k$ is updated in the same manner of \eqref{eqn:aoidevice} and \eqref{eqn:aoibs}.
We define $\bs{A}(t)\triangleq (\bs{A}_k(t))_{k\in\mathcal{K}}\in\mathcal{A}\triangleq\prod_{k\in\mathcal{K}}\mathcal{A}_k$, $\bs{h}(t)\triangleq (h_{k}(t))_{k\in\mathcal{K}}\in\mathcal{H}\triangleq\prod_{k\in\mathcal{K}}\mathcal{H}_k$, and $\bs{w}(t)=(\bs{w}_k(t))_{k\in\mathcal{K}}\in\mathcal{W}$ as the system AoI state, the system channel state, and the system control action at slot $t$, respectively.
Let $C_{s,k}$ and $C_{u,k}(h_k)$ be the sampling cost and the updating cost under channel state $h_k$ of IoT device $k$, respectively.
We assume that, the channel state processes $\{h_k(t)\} (k\in\mathcal{K})$ at the devices are mutually independent.
As in \cite{jiang2018can}, we consider that, in each slot, the multiple IoT devices cannot update their status packets concurrently; otherwise collisions occur and no status packets will be transmitted to the destination successfully. Thus, \emph{different from the case of a single IoT device, the updating process of the multiple IoT devices should be carefully scheduled to avoid such collisions}.
Mathematically, we have $ \sum_{k\in\mathcal{K}} u_k(t) \leq 1$, for all $t$.
% \begin{equation}
% \sum_{k\in\mathcal{K}} u_k(t) \leq 1, \forall t.
% \end{equation}
% Then, we define $\mathcal{W}\triangleq \{(s_k,u_k)_{k\in\mathcal{K}} | (s_k,u_k)\in\{0,1\}\times \{0,1\}, \forall k\in\mathcal{K}\text{~and~} \sum_{k\in\mathcal{K}} u_k \leq 1 \}$ as the feasible system control action space.
Then, we define $\mathcal{W}\triangleq \mathcal{S} \times\mathcal{U}$ as the feasible system control action space, where $\mathcal{S}\triangleq\{0,1\}^K$ and $\mathcal{U}\triangleq \{(u_k)_{k\in\mathcal{K}} | u_k\in\{0,1\} \forall k\in\mathcal{K}\text{~and~} \sum_{k\in\mathcal{K}} u_k \leq 1 \}$.
Note that, the proposed analytical framework and algorithm design can be readily extended to support the  orthogonal frequency division multiple access (OFDMA) mode, in which multiple IoT devices can update their status at the same time without collisions over different non-overlapping channels\cite{7962670}.

Similar to the single device case, given an observed system AoI state $\bs{A}$ and system channel state $\bs{h}$, the system control action $\bs{w}$ is derived as per the following policy.
\begin{definition}\label{definition:stationary_policy_MIoT}
A \emph{feasible stationary sampling and updating policy} $\pi=(\pi_{s},\pi_{u})$ is defined as a mapping from the system AoI state and the system channel state $(\bs{A},\bs{h})\in\mathcal{A}\times\mathcal{H}$ to the feasible system control action of the IoT devices $\bs{w}\in\mathcal{W}$, where $\pi_{s}(\bs{A},\bs{h})=\bs{s}$ and $\pi_{u}(\bs{A},\bs{h})=\bs{u}$.
\end{definition}

Under a given stationary unichain policy $\pi$, the average AoI at the destination and the average energy cost for each IoT device $k$ are respectively given by:
\begin{align}
   &\bar{A}_r(\pi)\triangleq\limsup_{T\to\infty}\frac{1}{T}\sum_{t=1}^T \sum_{k=1}^K \mathbb{E} \left[A_{r,k}(t)\right],\label{eqn:avg_aoi_miot}\\
   &\bar{C}_k(\pi)\triangleq\limsup_{T\to\infty}\frac{1}{T}\sum_{t=1}^T \mathbb{E} \left[C_k(\bs{w}_k(t))\right], \forall k\in\mathcal{K}, \label{eqn:avg_cost_miot}
\end{align}
where $C_k(\bs{w}_k(t))\triangleq s_k(t) C_{s,k} + u_k(t)C_{u,k}(h_k(t))$ and the expectation is taken with respect to the measure induced by the policy $\pi$.

We want to find the optimal \emph{feasible} sampling and updating policy that minimizes the average AoI at the destination, under an average energy cost
constraint for each IoT device, as follows:
\begin{subequations}\label{eqn:cmdp_miot}
\begin{align}
&\bar{A}_r^*\triangleq\min_{\pi}\bar{A}_r(\pi),\label{eqn:opt_Ar_miot}\\
&\text{s.t.~} \bar{C}_k(\pi) \leq C^{\textrm{max}}_k, \forall k\in\mathcal{K}.\label{eqn:constraint_miot}
\end{align}
\end{subequations}
To obtain the optimal policy $\pi^*$  in \eqref{eqn:cmdp_miot}, we again introduce the Lagrangian for a given vector of Lagrange multipliers $\bs{\lambda}\triangleq (\lambda_k)_{k\in\mathcal{K}}$, given by:
\begin{align}
   \bar{L}(\pi;\bs\lambda)\triangleq\limsup_{T\to\infty}\frac{1}{T}\sum_{t=1}^T \mathbb{E} \left[L(\bs{A}(t),\bs{h}(t),\bs{w}(t);\bs\lambda)\right],\label{eqn:avg_mdp_miot}
\end{align}
where $L(\bs{A}(t),\bs{h}(t),\bs{w}(t);\bs\lambda)\triangleq \sum_{k=1}^K \left(A_{r,k}(t) + \lambda_k (C_k(\bs{w}_k(t))-C^{\textrm{max}}_k)\right)$ is the Lagrange cost at slot $t$.
Then, the corresponding unconstrained MDP for a given $\bs{\lambda}$ will be:
\begin{align}
\bar{L}^*(\bs\lambda)\triangleq\min_{\pi}\bar{L}(\pi;\bs\lambda),\label{eqn:opt_L_miot}
\end{align}
% \end{subequations}
where  $\bar{L}^*(\bs\lambda)$ is the minimum average Lagrange cost achieved by the optimal policy $\pi^*_{\bs\lambda}$ for a given $\bs\lambda$.
The optimal average AoI at the destination in \eqref{eqn:opt_Ar_miot} is given by $\bar{A}_r^*=\max_{\bs\lambda} \bar{L}^*(\bs\lambda)$. In the following lemma, we summarize the solution to the unconstrained MDP in \eqref{eqn:opt_L_miot}.

\begin{lemma}\label{lemma:bellman_miot}
For any $\bs\lambda$, there exists $(\theta_{\bs\lambda},\{Q(\bs{A},\bs{h},\bs{u};\bs\lambda)\})$ satisfying:
% \begin{align}
%   \theta_{\lambda}&+V(\bs{A},h;\lambda)=\min_{\bs{w}\in\mathcal{W}}\Big\{L(\bs{A},h,\bs{w};\lambda)\nonumber\\&\hspace{1mm}+\sum_{h'\in\mathcal{H}}p_{\mathcal{H}}(h')V(\bs{A}',h';\lambda)\Big\},
%   ~\forall (\bs{A},h)\in\mathcal{A}\times\mathcal{H},\label{eqn:bellman}
% \end{align}
% \begin{align}
%   \theta_{\bs\lambda}+Q(\bs{A},\bs{h},\bs{u};\bs\lambda)&=\min_{\bs{s}\in\mathcal{S}}\Bigg\{L(\bs{A},\bs{h},\bs{w};\bs\lambda)\nonumber\\&\hspace{1mm}+\sum_{\bs{h}'\in\mathcal{H}}p_{\mathcal{H}}(\bs{h}') \min_{\bs{u}'\in\mathcal{U}}Q(\bs{A}',\bs{h}',\bs{u}';\bs\lambda)\Bigg\},
%   ~\forall (\bs{A},\bs{h},\bs{u})\in\mathcal{A}\times\mathcal{H}\times\mathcal{U},\label{eqn:bellman_miot}
% \end{align}
\begin{align}
  \theta_{\bs\lambda}+Q(\bs{A},\bs{h},\bs{u};\bs\lambda)=\min_{\bs{s}\in\mathcal{S}}\Big\{L(\bs{A},\bs{h},\bs{w};\bs\lambda)+\sum_{\bs{h}'\in\mathcal{H}}p_{\mathcal{H}}(\bs{h}') \min_{\bs{u}'\in\mathcal{U}}Q(\bs{A}',\bs{h}',\bs{u}';\bs\lambda)\Big\},\label{eqn:bellman_miot}
\end{align}
where $\bs{A}'$ satisfies the AoI dynamics in \eqref{eqn:aoidevice} and \eqref{eqn:aoibs} for each IoT device,
$\theta_{\bs\lambda}=\bar{L}^*(\bs\lambda)$ is the optimal value to \eqref{eqn:opt_L_miot} for all initial state $(\bs{A}(1),\bs{h}(1),\bs{u}(1))$, and $Q(\cdot)$ is the Q-factor \textcolor{black}{which is a mapping from $(\bs{A},\bs{h},\bs{u})$ to real values.}
Moreover, for a given $\bs\lambda$, the optimal policy achieving the optimal value $\bar{L}^*(\bs\lambda)$ is given by $\pi^*_{\bs\lambda}(\bs{A},\bs{h})= (\pi_{\bs\lambda,s}^*(\bs{A},\bs{h}),\pi_{\bs\lambda,u}^*(\bs{A},\bs{h}))$, where $\pi_{\bs\lambda,s}^*(\bs{A},\bs{h})$ attains the minimum of the right-hand side of \eqref{eqn:bellman_miot} and $\pi_{\bs\lambda,u}^*(\bs{A},\bs{h})=\arg\min_{\bs{u}\in\mathcal{U}} Q(\bs{A},\bs{h},\bs{u};\bs\lambda)$.
\end{lemma}

\begin{IEEEproof}
See Appendix E.
\end{IEEEproof}

From Lemma~\ref{lemma:bellman_miot}, we can see that the optimal sampling and updating action depends on the Q-factor $Q(\bs{A},\bs{h},\bs{u};\bs\lambda)$ and the $K$ Lagrange multipliers. For a given $\bs\lambda$,  obtaining the Q-factor $Q(\cdot)$ requires solving the Bellman equation in \eqref{eqn:bellman_miot}, which suffers from the curse of the dimensionality due  to the exponential growth of the cardinality of the system state space ($|\mathcal{A}\times\mathcal{H}|=\prod_{k=1}^K A_{l,k}^\textrm{max}A_{r,k}^\textrm{max}|\mathcal{H}_k|$).
Even if we could obtain the optimal Q-factors by solving \eqref{eqn:bellman_miot}, the derived control will be centralized thus requiring a knowledge of the system AoI states and channel states at each slot by the destination node, which is highly undesirable. Moreover, the optimal policy of the CMDP in \eqref{eqn:cmdp_miot} is a randomized stationary policy with a degree of randomization no greater than $K$ \cite{cmdp_multiple}, and, thus, may not be very suitable for practical implementations.
Note that, since we need to jointly control the sampling and updating processes, our problem cannot be cast into a restless multi-armed bandit problem (RMAB)\cite{gittins2011multi} as is often done in the literature\footnote{\textcolor{black}{In general, RMAB only works for the problem with only one type of control actions.}}, thus rendering the existing low-complexity solutions (e.g.,\cite{kadota2018scheduling,HsuISIT}, and \cite{jiang2018can}) not applicable.
% \textcolor{black}{Note that, since we are jointly controlling the sampling and updating processes, our problem cannot be cast into a restless multi-armed bandit problem (RMAB)\cite{gittins2011multi}, and, thus, the existing low-complexity solutions (e.g.,\cite{kadota2018scheduling,jiang2018can,HsuISIT}) based on RMAB cannot be applied to our considered problem.}
Therefore, we next introduce a novel semi-distributed low-complexity algorithm to obtain a deterministic suboptimal sampling and updating policy.

\subsection{Algorithm Design}
In this subsection, we first approximate the Q-factor $Q(\bs{A},\bs{h},\bs{u};\bs\lambda)$ by the sum of the per-device Q-factor $Q_k(\bs{A}_k,h_k,u_k;\lambda_k)$.
% , i.e., $Q(\bs{A},\bs{h},\bs{u};\bs\lambda)\thickapprox\sum_{k=1}^K Q_k(\bs{A}_k,h_k,u_k;\lambda_k)$.
Based on the approximated Q-factor, we propose a semi-distributed sampling and updating policy, inspired by \cite{5462936}. Then, we develop an online learning algorithm that enables each device to determine its per-device Q-factor and the associated Lagrange multiplier based on the observation of its AoI and channel states. Finally, we show that the proposed semi-distributed learning algorithm converges to the proposed suboptimal policy.
 % Based on the approximate Q-factor, we propose an auction-based distributed sampling and updating policy, inspired by \cite{surveyit,5462936}. Then, we propose a localized online learning algorithm that enables each IoT device to determine its per-device Q-factor and the associated Lagrange multiplier. We prove that the proposed auction-based distributed online learning algorithm converges almost surely.

\subsubsection{Semi-Distributed Sampling and Updating Control}
\textcolor{black}{To reduce the complexity for obtaining the optimal Q-factor}, we  adopt the linear approximation architecture \cite{bertsekas} to approximate the Q-factor in \eqref{eqn:bellman_miot} by the sum of the per-device Q-factor $Q_k(\bs{A}_k,h_k,\allowbreak u_k;\lambda_k)$:
\begin{equation}
Q(\bs{A},\bs{h},\bs{u};\bs\lambda)\thickapprox\sum_{k=1}^K Q_k(\bs{A}_k,h_k,u_k;\lambda_k),\label{eqn:Q_factor_approx}
\end{equation}
where $Q_k(\bs{A}_k,h_k,u_k;\lambda_k)$ satisfies the following per-device Q-factor fixed point equation of each IoT device $k$ for each given $\lambda_k$:
\begin{align}
  \theta_k+&Q_k(\bs{A}_k,h_k,u_k;\lambda_k)  =\min_{s_k\in\{0,1\}}\Bigg\{L_k(\bs{A}_k,h_k,s_k,u_k;\lambda_k)\nonumber\\&\hspace{5mm}+\sum_{h_k'\in\mathcal{H}_k}p_{\mathcal{H}_k}(h_k') \min_{u_k'\in\{0,1\}} Q_k(\bs{A}_k',h_k',u_k';\lambda_k)\Bigg\},
  ~\forall (\bs{A}_k,h_k,u_k)\in\mathcal{A}_k\times\mathcal{H}_k\times\{0,1\}.\label{eqn:bellman_miot_perdevice}
\end{align}
Here, $L_k(\bs{A}_k,h_k,s_k,u_k;\lambda_k)=A_{r,k} + \lambda_k (C_k(\bs{w}_k)-C^{\textrm{max}}_k)$ is the per-device Lagrange cost for IoT device $k$.
Then, according to  Lemma~\ref{lemma:bellman_miot}, the destination node determines the updating control policy of all IoT devices based on  the linear approximation in \eqref{eqn:Q_factor_approx}, given by:
\begin{equation}
\hat{\pi}_{\bs\lambda,u}^*(\bs{A},\bs{h}) = \arg\min_{\bs{u}\in\mathcal{U}} \sum_{k=1}^K Q_k(\bs{A}_k,h_k,u_k;\lambda_k). \label{eqn:auction_updating_policy}
\end{equation}
% This can be determined by using a per-stage auction, where the $K$ IoT devices, i.e., the bidders, submit the bid to the destination node, i.e., the auctioneer.
\textcolor{black}{Problem \eqref{eqn:auction_updating_policy} can be solved by a brute-force search with complexity of $O(|\mathcal{U}|)$.}
\textcolor{black}{In particular, each IoT device $k$ observes its AoI state $\bs{A}_k$ and channel state $h_k$ and reports its current per-device Q-factor $\{Q_k(\bs{A}_k,h_k,u_k;\lambda_k), u_k=0,1\}$ to the destination node. Then, the destination node determines the system updating action $\hat{\bs{u}}^*=\hat{\pi}_{\bs\lambda,u}^*(\bs{A},\bs{h})$  according to \eqref{eqn:auction_updating_policy} and broadcasts the updating action $\hat{\bs{u}}^*=(\hat{u}_k^*)_{k\in\mathcal{K}}$ to the $K$ IoT devices.}
Based on the local observation of $\bs{A}_k$ and $h_k$, as well as the updating action $\hat{u}_k^*$, each IoT device $k$ decides its sampling action $\hat{s}_k^*$, which minimizes the right-hand side of \eqref{eqn:bellman_miot_perdevice}:
\begin{equation}
% s_k^*=\arg\min_{s_k\in\{0,1\}}\left\{L_k(\bs{A}_k,h_k,s_k,u_k^*;\lambda_k)+\sum_{h_k'\in\mathcal{H}_k}p_{\mathcal{H}_k}(h_k') \min_{u_k'\in\{0,1\}} Q_k(\bs{A}_k',h_k',u_k';\lambda_k)\right\}.
\hat{s}_k^*=\arg\min_{s_k\in\{0,1\}}\bigg\{L_k(\bs{A}_k,h_k,s_k,\hat{u}_k^*;\lambda_k)+\sum_{h_k'\in\mathcal{H}_k}p_{\mathcal{H}_k}(h_k') \min_{u_k'\in\{0,1\}} Q_k(\bs{A}_k',h_k',u_k';\lambda_k)\bigg\}.\label{eqn:auction_sampling_policy}
\end{equation}

Note that, to obtain the proposed suboptimal policy $\hat{\pi}^*$ \textcolor{black}{in \eqref{eqn:auction_updating_policy} and \eqref{eqn:auction_sampling_policy}}, we need to compute $Q_k(\bs{A}_k,h_k,u_k;\lambda_k)$ by solving \eqref{eqn:bellman_miot_perdevice} for all IoT devices, which is a total of $O(\sum_{k=1}^K A_{l,k}^\textrm{max}A_{r,k}^\textrm{max}|\mathcal{H}_k|)$  values.
However, to obtain the optimal policy $\pi^*$, computing $Q(\bs{A},\bs{h},\bs{u};\bs\lambda)$ by solving \eqref{eqn:bellman_miot} requires a total of $O(\prod_{k=1}^K A_{l,k}^\textrm{max}A_{r,k}^\textrm{max}|\mathcal{H}_k|)$ values.
Therefore, the complexity of the proposed suboptimal policy decreases from exponential with $K$ to linear with $K$.
\textcolor{black}{\footnote{\textcolor{black}{The approximation error analysis of the linear approximation in \eqref{eqn:Q_factor_approx} (which is a feature-based method) remains an open problem for CMDPs. Thus, we only provide numerical comparisons to illustrate its performance in the simulations.}}}

\subsubsection{Online Stochastic Learning and Convergence Analysis}
We observe that the proposed semi-distributed policy $\hat{\pi}^*$ requires the knowledge of the per-device Q-factor and the associated Lagrange multiplier, which is challenging to obtain.
Thus, we propose an online learning algorithm to estimate $Q_k(\bs{A}_k,h_k,u_k;\lambda_k)$  and $\lambda_k$ at each IoT device $k$.

For IoT device $k$, based on the locally observed AoI state $\bs{A}_k(t)$, channel state $h_k(t)$, the updating action $\hat{u}_k^*(t)$ from the destination node, and the sampling action $\hat{s}_k^*(t)$, the per-device Q-factor and the Lagrange multiplier are respectively updated according to
\begin{align}
&Q_k^{t+1}(\bs{A}_k,h_k,u_k;\lambda_k^t)
= Q_k^{t}(\bs{A}_k,h_k,u_k;\lambda_k^t)  + \epsilon^{\upsilon_k^t(\bs{A}_k,h_k,u_k)}_{q,k} \Big(F_k(\bs{A}_k,h_k,u_k,\hat{s}_k^*(t);\lambda_k^t)\label{eqn:update_Q_factor}\\
&-F_k(\bs{A}_k^{r},h_k^{r},u_k^{r},\hat{s}_k^*(t^{r});\lambda_k^t) - Q_k^{t}(\bs{A}_k,h_k,\hat{u}_k^*;\lambda_k^t)\Big) \mathbbm{1}\left((\bs{A}_k(t),h_k(t),\hat{u}_k^*(t))=(\bs{A}_k,h_k,u_k)\right),\nonumber\\
&\lambda_k^{t+1} = [\lambda_k^{t}+\epsilon_{\lambda,k}^t(C_k(\hat{\bs{w}}_k(t))-C^{\textrm{max}}_k)]^ +,\label{eqn:update_lambda}
\end{align}
where $\mathbbm{1}(\cdot)$ is the indicator function, $\upsilon_k^\tau(\bs{A}_k,h_k,u_k)\triangleq \allowbreak \sum_{\tau=1}^t \mathbbm{1}((\bs{A}_k(\tau),h_k(\tau),\hat{u}_k^*(\tau))\allowbreak=(\bs{A}_k,h_k,u_k))$ is the number of updates of the state-action pair $(\bs{A}_k,h_k,u_k)$ till $t$,
$F_k(\bs{A}_k,h_k,u_k,s_k;\lambda_k^t)\triangleq L_k(\bs{A}_k,h_k,s_k,u_k;\lambda_k^t)+\sum_{h_k'\in\mathcal{H}_k}p_{\mathcal{H}_k}(h_k') \min_{u_k'\in\{0,1\}} Q_k^t(\bs{A}_k',h_k',u_k';\lambda_k^t)$ with $\bs{A}_k'$ and $\bs{A}_k$ satisfying the relations in \eqref{eqn:aoidevice} and \eqref{eqn:aoibs},
$(\bs{A}_k^{r},h_k^{r},u_k^{r})$ is some fixed reference state-action pair,  $t^{r}\triangleq\sup\{t|(\bs{A}_k(t),h_k(t),\hat{u}_k^*(t))=(\bs{A}_k^{r},h_k^{r},u_k^{r})\}$, and $[x]^+=\max\{x,0\}$. $\{\epsilon^t_{q,k}\}$ and $\{\epsilon_{\lambda,k}^t\}$ are the sequences of step sizes satisfying:
\begin{align}
&\sum_t \epsilon^t_{q,k} = \infty,~\epsilon^t_{q,k}>0, \lim_{t\to \infty} \epsilon^t_{q,k} = 0, \sum_t \epsilon_{\lambda,k}^t  = \infty,~\epsilon_{\lambda,k}^t>0, \lim_{t\to\infty} \epsilon_{\lambda,k}^t = 0, \nonumber\\
&\sum_t ((\epsilon^t_{q,k})^2+(\epsilon_{\lambda,k}^t)^2) < \infty,~\text{and}~\lim_{t\to\infty} \frac{\epsilon_{\lambda,k}^t}{\epsilon^t_{q,k}} = 0. \label{eqn:stepsize}
\end{align}
\textcolor{black}{Here, \eqref{eqn:update_Q_factor} is formulated following the asynchronous relative value Q-learning algorithm\cite{abounadi2001learning}.}

% According to \eqref{eqn:update_Q_factor} and \eqref{eqn:update_lambda}, the proposed online learning algorithm requires only the local AoI and channel states, and  both the per-device Q-factor and the Lagrange multiplier are updated simultaneously at each IoT device.
From \eqref{eqn:update_Q_factor} and \eqref{eqn:update_lambda}, to implement the proposed online learning algorithm at each IoT device, we only need the local AoI and channel states, as well as the updating control action from the destination.
The proposed algorithm is illustrated in Algorithm~\ref{alg:learning}.
\textcolor{black}{It can be seen that the proposed algorithm is essentially a grant-based uplink transmission protocol (see examples in\cite{7876968,hasan2013random}), which involves the exchange of messages between the IoT devices and the destination node. However, this will not incur any notable overhead, because in each slot,  each IoT device needs to only transmit a few bits to exchange its  per-device Q-factor value with the destination. }
Note that, we need to update both the per-device Q-factors and the Lagrange multipliers simultaneously. 
Thus, conventional value iteration and policy iteration algorithms\cite{bertsekas}, and the Q-learning algorithm under which the Lagrange multipliers are determined offline \cite{djonin2007q} are not applicable to our case.

\begin{algorithm}[!t]
% \small
\caption{Semi-Distributed Sampling and Updating Learning Algorithm.}
\label{alg:learning}
\begin{algorithmic}[1]
\State \textbf{Initialization:} Set $t=1$. Each IoT device initializes its per-device Q-factor $Q_k^t(\cdot)$ and Lagrange multiplier $\lambda_k^t$.

\State \textbf{Updating control \textcolor{black}{at the destination}:} At slot $t$, each IoT device $k$ reports $\{Q_k(\bs{A}_k(t),h_k(t),u_k;\lambda_k(t)), u_k=0,1\}$ to the destination node. Then, the destination node determines the system updating action according to \eqref{eqn:auction_updating_policy}
and broadcast the updating action $\hat{\bs{u}}^*(t)=(\hat{u}_k^*(t))_{k\in\mathcal{K}}$ to the $K$ IoT devices. \label{step:auction}

\State \textbf{Sampling control \textcolor{black}{at each IoT device}:} Based on the updating action $u_k^*(t)$, each IoT device $k$ decides its sampling action $\hat{s}_k^*(t)$  according to \eqref{eqn:auction_sampling_policy}.

\State \textbf{Per-device Q-factor and Lagrange multiplier update \textcolor{black}{at each IoT device}:} Based on the current observations $\bs{A}_k(t)$ and $h_k(t)$, each IoT device $k$ updates the per-device Q-factor $Q_k^{t+1}(\bs{A}_k,h_k,u_k;\lambda_k^t)$ and $\lambda_k^{t+1}$ according to \eqref{eqn:update_Q_factor} and \eqref{eqn:update_lambda}, respectively. \label{step:update}

\State  Set $t\leftarrow t+1$ and go to Step~\ref{step:auction} until the convergence of $Q_k^t(\cdot)$ and $\lambda_k^t$.
	
\end{algorithmic}
\end{algorithm}

Now, we show the almost-sure convergence of Algorithm~\ref{alg:learning}.
% Note that, different from the conventional online learning, where the centralized control is determined entirely on the value function or Q-factor update\cite{bertsekas}, for our proposed algorithm, the control action of each IoT device is determined by a per-stage auction. During the iterative updates, the per-device Q-factors, the Lagrange multipliers, and the control actions are changed dynamically, which makes the local per-factor Q-factor update in \eqref{eqn:update_Q_factor} not a contraction mapping. Therefore, the conventional convergence analysis using contraction mapping argument cannot be directly applied to our proposed distributed learning algorithm.
From \eqref{eqn:stepsize}, we can see that the per-device Q-factor and the Lagrange multiplier are updated concurrently, albeit over two different timescales\cite{stochasticlearning}. During the update of the per-device Q-factor (timescale I), we have $\lambda_k^{t+1}-\lambda_k^t=O(\epsilon_{\lambda,k}^t)=o(\epsilon_{q,k}^t)$, and thus, $\lambda_k^t$ can be seen as quasi-static \cite{stochasticlearning} when updating $Q_k^{t}(\bs{A}_k,h_k,u_k;\lambda_k^t)$ in \eqref{eqn:update_Q_factor}. We first have the following lemma on the convergence of the per-device Q-factor learning over timescale I.
\begin{lemma}\label{lemma:convergence_of_Q_factor}
For step sizes $\{\epsilon^t_{q,k}\}$ and $\{\epsilon_{\lambda,k}^t\}$ satisfying the conditions in \eqref{eqn:stepsize}, the update of the per-device Q-factor at each IoT device $k$ converges almost surely \textcolor{black}{to the solution of the fixed point equation in\eqref{eqn:bellman_miot_perdevice},} under any initial per-device Q-factor $Q_k^1(\cdot)$ and the Lagrange multiplier vector $\bs{\lambda}$, i.e., $\lim_{t\to\infty} Q_k^t(\bs{A}_k,h_k,u_k;\lambda_k) = \bs{Q}_k^{\infty}(\bs{A}_k,h_k,u_k;\lambda_k), \text{~a.s.~}, \forall \bs{A}_k,h_k,u_k,k.$
% \begin{equation}
% \lim_{t\to\infty} Q_k^t(\bs{A}_k,h_k,u_k;\lambda_k) = \bs{Q}_k^{\infty}(\bs{A}_k,h_k,u_k;\lambda_k), \text{~a.s.~}, \forall \bs{A}_k,h_k,u_k,k.
% \end{equation}
\end{lemma}

\begin{IEEEproof}
See Appendix F.
\end{IEEEproof}

During the update of the Lagrange multiplier (timescale II) in \eqref{eqn:update_lambda}, the per-device Q-factor can be seen as nearly equilibrated\cite{stochasticlearning}. Then, we have the following convergence result.
\begin{lemma}\label{lemma:convergence_of_lambda}
The update of the vector of the Lagrange multipliers $\bs{\lambda}$ converges almost surely, i.e., $\lim_{t\to\infty} \bs{\lambda}^t = \bs{\lambda^{\infty}}$ a.s.,
% \begin{equation}
% \lim_{t\to\infty} \bs{\lambda}^t = \bs{\lambda^{\infty}}, \text{~a.s.~},
% \end{equation}
where the policy under $\bs{\lambda^{\infty}}$ satisfies the constraints in \eqref{eqn:constraint_miot}.
\end{lemma}

\begin{IEEEproof}
See Appendix G.
\end{IEEEproof}

Based on Lemma~\ref{lemma:convergence_of_Q_factor} and Lemma~\ref{lemma:convergence_of_lambda}, we summarize the convergence of the proposed semi-distributed online sampling and updating algorithm in Algorithm~\ref{alg:learning} in the following Theorem.
\begin{theorem}
For step sizes $\{\epsilon^t_{q,k}\}$ and $\{\epsilon_{\lambda,k}^t\}$ satisfying the conditions in \eqref{eqn:stepsize}, the iterations of the per-device Q-factor and the Lagrange multipliers in Algorithm~\ref{alg:learning} converge w.p. 1, i.e., $(\bs{Q}^t_k,\lambda_k^t)\rightarrow (\bs{Q}^{\infty}_k,\lambda_k^{\infty})$ almost surely, for each IoT device $k$, where $(\bs{Q}^{\infty}_k,\lambda_k^{\infty})$ satisfies the fixed-point equation in \eqref{eqn:bellman_miot_perdevice} and the sampling and updating policy under $(\bs{Q}^{\infty}_k,\lambda_k^{\infty})$  satisfies the average energy cost constraints in \eqref{eqn:constraint_miot}.
\end{theorem}

In a nutshell,  we have proposed a low-complexity semi-distributed learning algorithm to find a suboptimal sampling and updating policy so as to minimize the average AoI at the destination for the case of multiple IoT devices.
 The proposed semi-distributed learning algorithm can be implemented at each device, requiring only the local knowledge and simple signaling from the destination, and, thus, is highly desirable for practical implementations.

\section{Simulation Results and Analysis}
% In this section, we provide numerical examples to evaluate the performance of the proposed solutions for the cases of the single and multiple IoT devices.
\subsection{Case of A Single IoT Device}\label{sec:simu_single}
We first illustrate the structural properties of the optimal sampling and updating policy for the single IoT device case.
\textcolor{black}{In the simulations, we set $\mathcal{H}=\{0.0131, 0.0418, 0.0753, 0.1157,0.1661,\allowbreak 0.2343,0.3407, 0.6200\}$ and the corresponding probabilities are $p_{\mathcal{H}} = [1,1,2,3,3,2,1,1]/14$ \cite{6585729}.
Similar to  \cite{6585729}, we assume that the updating cost is $C_u(h)=C_u/h$, where $C_u=0.2$.
For the sampling cost, we adopt the local-computing model in \cite{7762913} and assume that $C_s=0.2$.
 % Unless otherwise stated, the sampling and updating costs are assumed to be $C_s=0.2$ and $C_u(h)=C_u/h$, respectively, where $C_u=0.2$, similar to  \cite{7762913} and \cite{6585729}.
 }
We set the upper limits of the AoI at the device and the AoI at the destination  $\hat{A}_l$ and $\hat{A}_r$ be $10$.

\begin{figure}[t]
\begin{minipage}[h]{.475\linewidth}
\centering
        \includegraphics[scale=0.5]{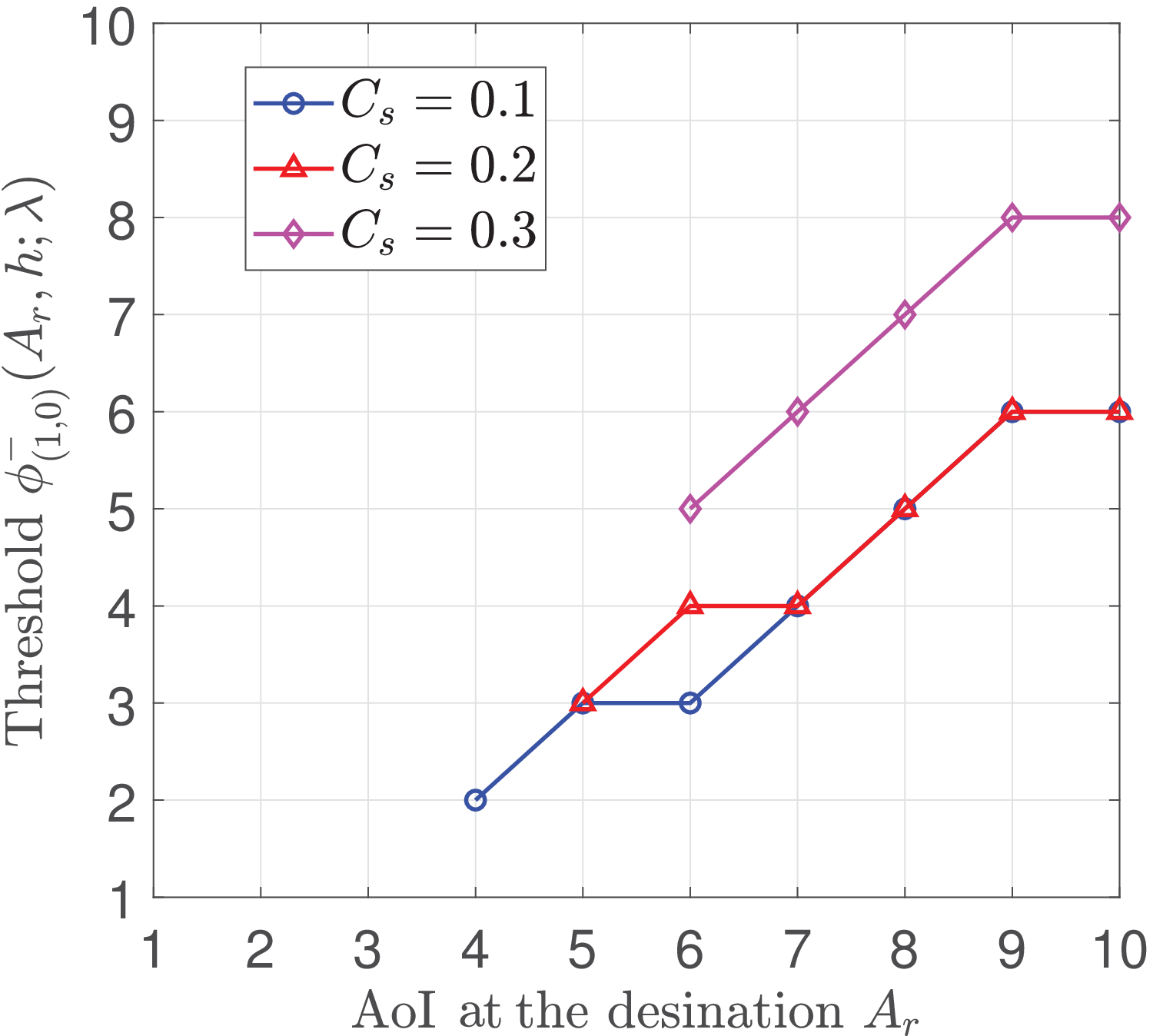}
\subcaption{}\label{fig:Cs}
\end{minipage}%
\begin{minipage}[h]{.475\linewidth}
\centering
        \includegraphics[scale=0.5]{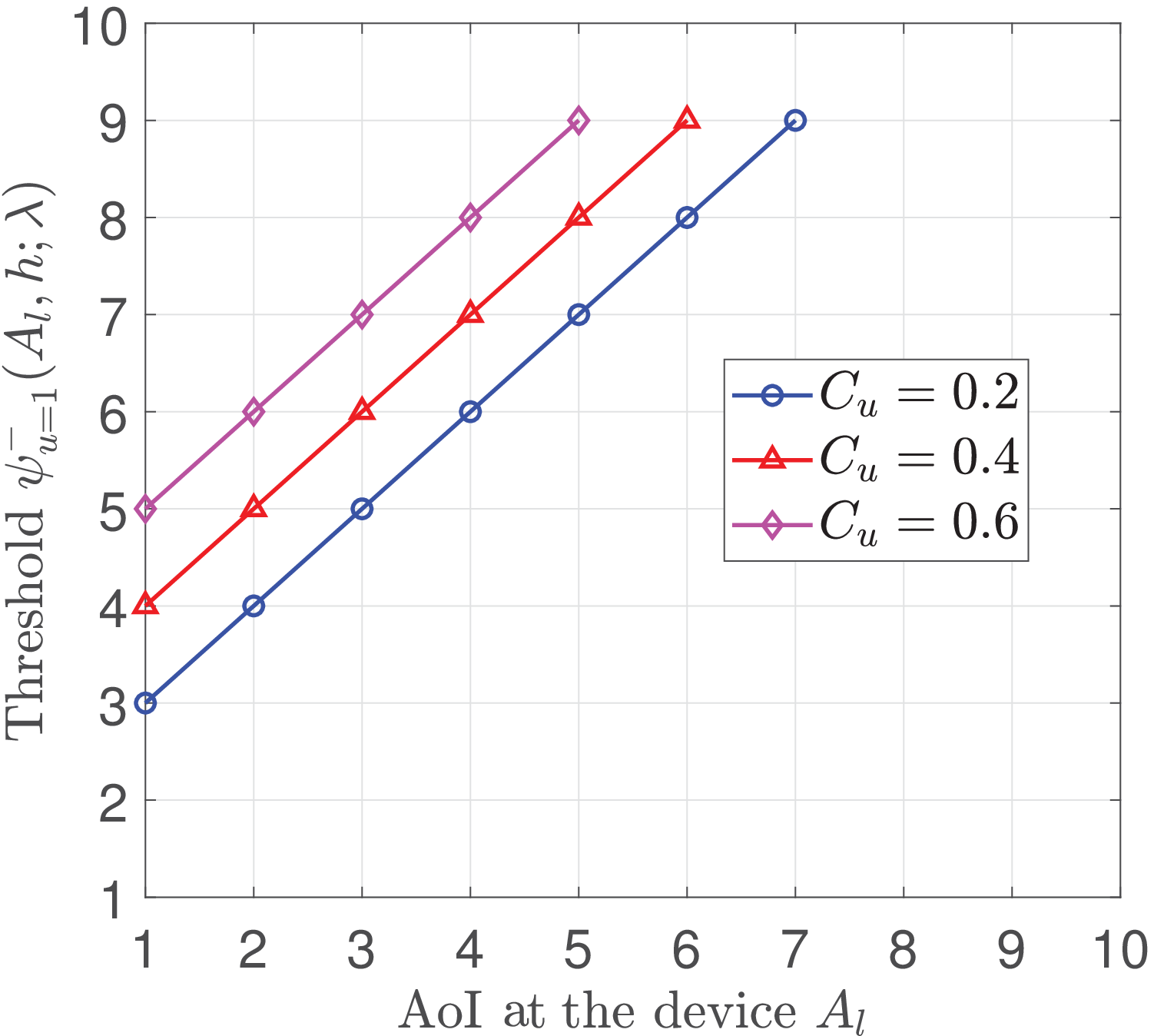}
\subcaption{}\label{fig:Cu}
\end{minipage}
% \vspace{-0.3cm}
\caption{Impacts of the sampling and updating costs on structures of the optimal policy $\pi_{\lambda}^*$ for a given $\lambda$ in the single IoT device case. $\lambda=0.1$.  (a)  Sampling cost. $h=0.0418$.  (b) Updating cost. $h=0.1157$.}\label{fig:structure_of_threshold}
% \vspace{-0.7cm}
\end{figure}

Fig.~\ref{fig:structure_of_threshold} illustrates the effects of the sampling and updating costs on the structural properties of the optimal policy $\pi_{\lambda}^*$ for a given $\lambda$ as shown in Theorem~\ref{theorem:optimal}. In particular, Fig.~\ref{fig:Cs} shows the relationship between the threshold $\phi_{(1,0)}^-(A_r,h;\lambda)$ of choosing action $(1,0)$  and $A_r$ under different sampling costs $C_s$, for given $h$ and $\lambda$. From Fig.~\ref{fig:Cs}, we can see, $\phi_{(1,0)}^-(A_r,h;\lambda)$ is non-decreasing with $C_s$. This indicates that the IoT device is unlikely to sample the physical process, if the sampling cost is high. Fig.~\ref{fig:Cu} shows the relationship between $\psi_{u=1}^- (A_l,h;\lambda) \triangleq \min\{\psi_{(0,1)}^- (A_l,h;\lambda),\psi_{(1,1)}^- (A_l,h;\lambda)\}$ and $A_l$ under different updating costs $C_u$, for given $h$ and $\lambda$. According to Theorem~\ref{theorem:optimal}, if $A_r\geq \psi_{u=1}^- (A_l,h;\lambda)$, then the optimal updating action is $u=1$, as $\pi^*_{\lambda}(\bs{A},h)=(0,1)~\text{or}~(1,1)$. We observe that, $\phi_{(1,0)}^-(A_r,h;\lambda)$ is  non-decreasing with $C_u$. This indicates that the IoT device is not willing to send the status packet to the destination, if the updating cost is high.

\begin{figure}[t]
\begin{minipage}[h]{.475\linewidth}
\centering
        \includegraphics[scale=0.5]{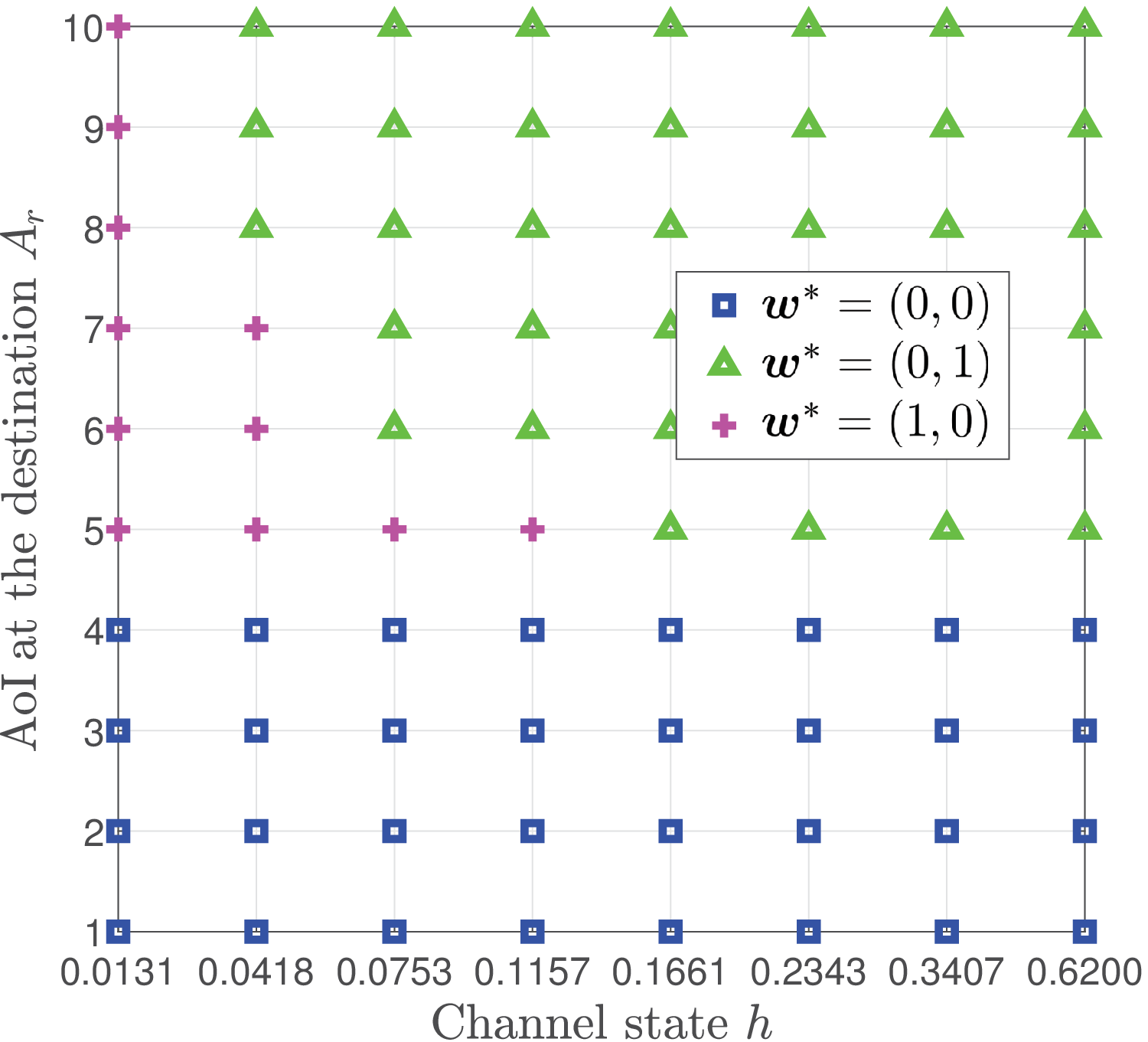}
\subcaption{}\label{fig:action01}
\end{minipage}%
\begin{minipage}[h]{.475\linewidth}
\centering
        \includegraphics[scale=0.5]{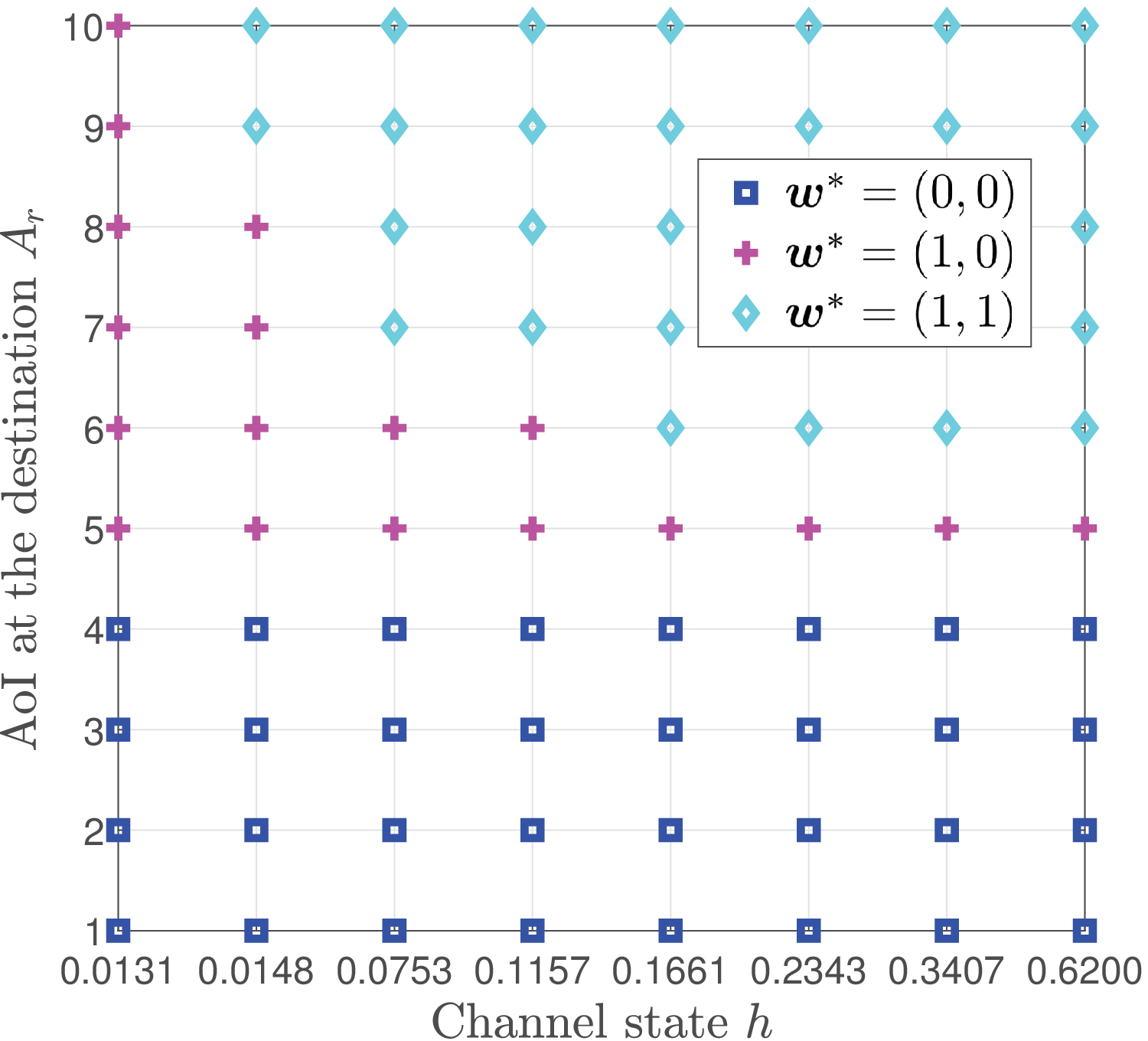}
\subcaption{}\label{fig:action11}
\end{minipage}
% \vspace{-0.4cm}
\caption{Structure of the optimal policy $\pi_{\lambda}^*$ for given $A_l$ and $\lambda$. $\lambda=0.1$. (a) $A_l=4$. (b) $A_l=5$.}\label{fig:structure_of_channel}
% \vspace{-0.1cm}
\end{figure}

Fig.~\ref{fig:structure_of_channel} shows the structure of the optimal sampling and updating policy $\pi_{\lambda}^*$ for given $A_l$ and $\lambda$.
From Fig.~\ref{fig:action01} and Fig.~\ref{fig:action11}, we can see that the scheduling of action $(0,1)$ or action $(1,1)$ is threshold-based with respect to the channel state $h$.
In particular, Fig.~\ref{fig:structure_of_channel} shows that,  if the channel state is poor, it is not efficient for the IoT device to send the status packet to the destination, as a high updating cost will be incurred. Therefore, the optimal policy can fully exploit the random nature of the wireless channel by seizing good transmission opportunities to optimize the system performance. We also notice that the optimal actions $(0,1)$ and $(1,1)$ do not concurrently appear in the whole state space of $(A_r,h)$, under a given $A_l$.
\textcolor{black}{This is due to the fact that the decisions of choosing $(0,1)$ or $(1,1)$ are threshold-based with respect to $A_r$ and $h$, as seen in the upper right corners of Fig.~\ref{fig:action01} and Fig.~\ref{fig:action11}.}
% This is due to the threshold-based updating scheduling of $(0,1)$ and $(1,1)$ with respect to $A_r$ and $h$, \textcolor{black}{as seen in the upper right corners of Fig.~\ref{fig:action01} and Fig.~\ref{fig:action11}.}

\begin{figure}[!t]
\begin{centering}
\includegraphics[scale=.6]{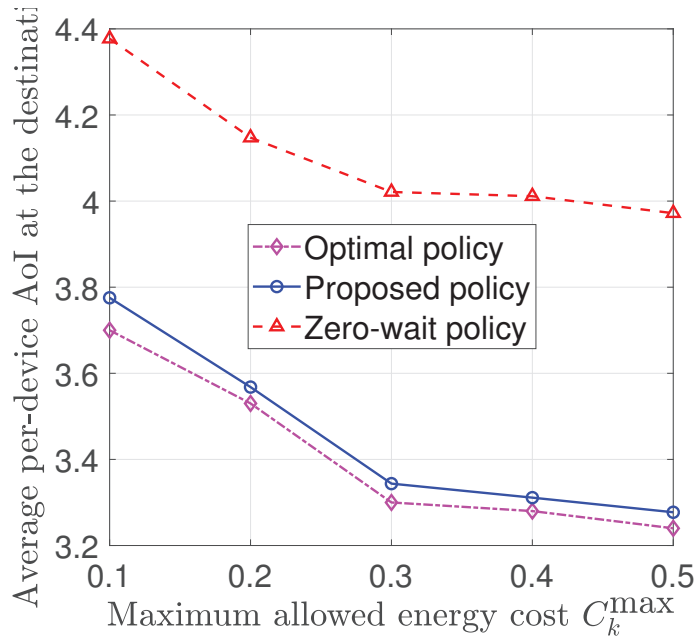}
% \vspace{-0.2cm}
 \caption{Performance comparison among the optimal policy, the proposed semi-distributed policy, and the zero-wait baseline policy. $K=2$, $A_{l,k}^\textrm{max}=A_{r,k}^\textrm{max}=20, \forall k=1,2$.}\label{fig:performance_op_sub}
\end{centering}
% \vspace{-0.7cm}
\end{figure}

\begin{figure}[!t]
\begin{minipage}[h]{.475\linewidth}
\centering
        \includegraphics[scale=0.5]{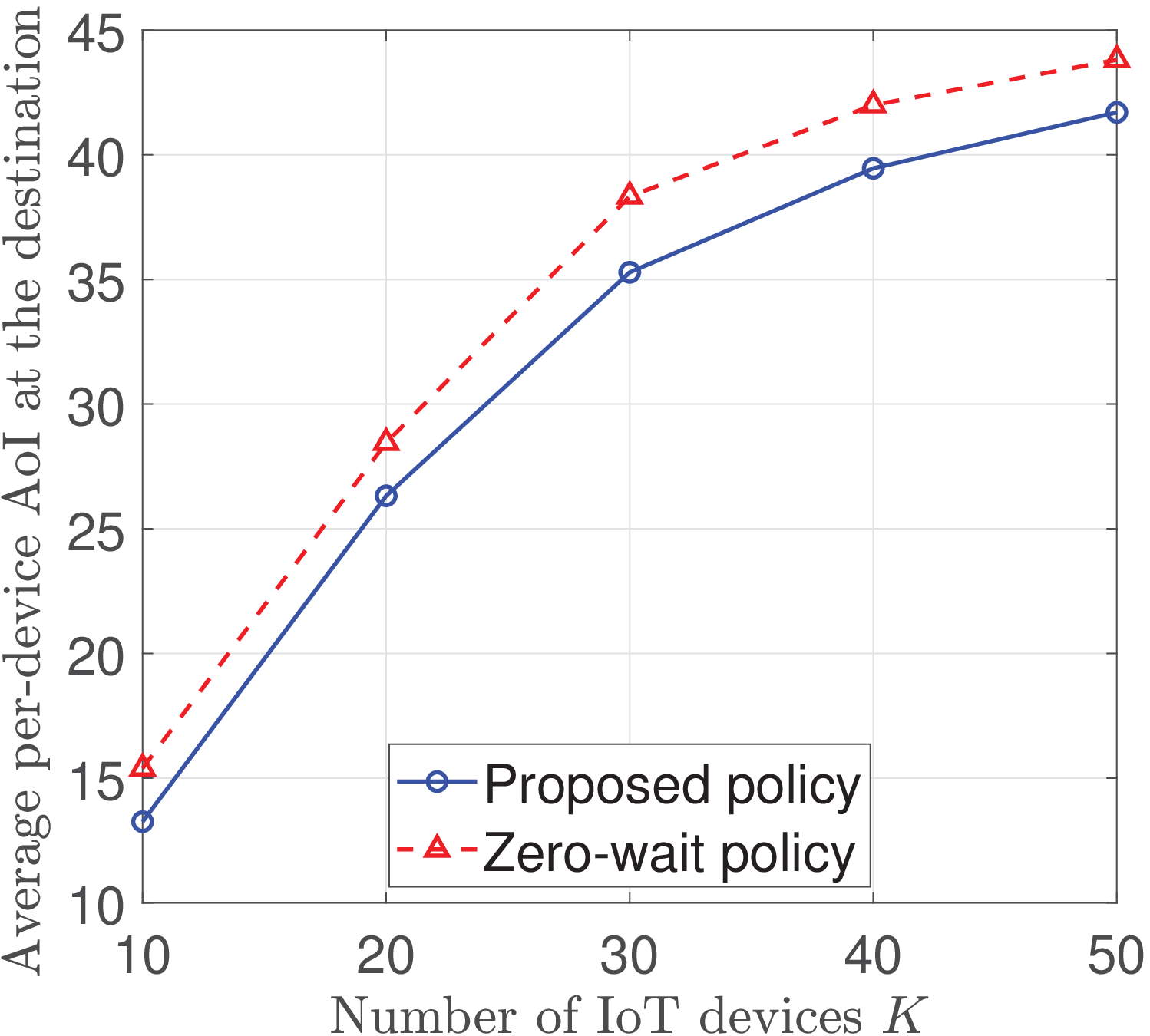}
\subcaption{}\label{fig:aoi_vs_numusers}
\end{minipage}%
\begin{minipage}[h]{.475\linewidth}
\centering
        \includegraphics[scale=0.5]{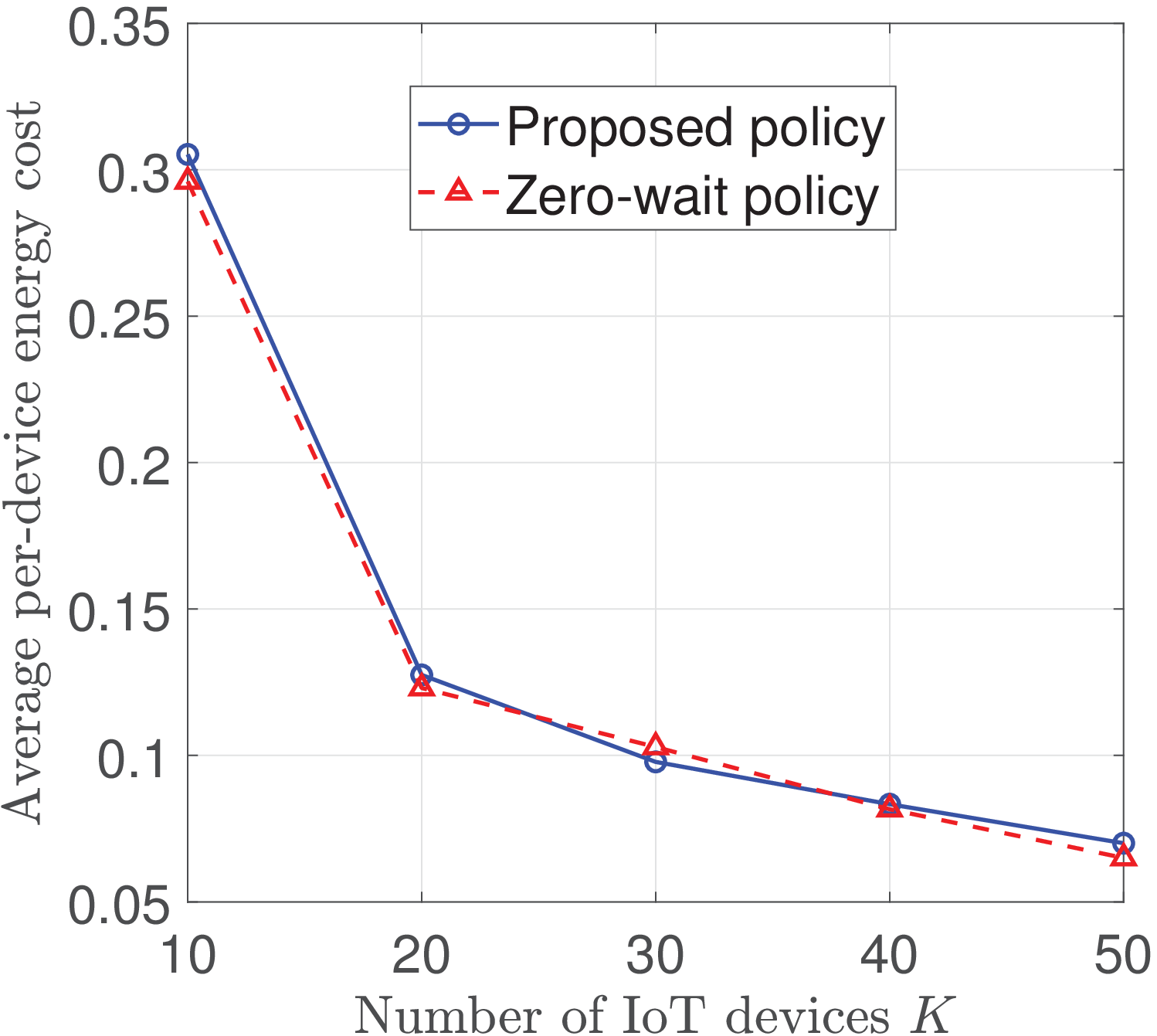}
\subcaption{}\label{fig:cost_vs_numusers}
\end{minipage}
% \vspace{-0.3cm}
\caption{Performance comparison between the proposed semi-distributed policy and the zero-wait baseline policy. $C^{\textrm{max}}_k=0.3, \forall k$.
\textcolor{black}{$A_{l,k}^\textrm{max}=A_{r,k}^\textrm{max}=100, \forall k$.}
 (a) Average  per-device AoI at the destination.  (b) Average per-device energy cost.}\label{fig:aoi_cost_numusers}
% \vspace{-0.7cm}
\end{figure}

\subsection{Case of Multiple IoT Devices}
Next, we evaluate the performance of the proposed semi-distributed online sampling and updating policy in Algorithm~\ref{alg:learning}.
 The system parameters are analogous to those for the single IoT device case.
\textcolor{black}{For each device $k$, the sampling cost $C_{s,k}$ is randomly selected from $[0.2,0.3]$ and the updating cost is $C_{u,k}(h_k) = C_{u,k}/h_k$, where $C_{u,k}$ is randomly selected from $[0.2,0.3]$.}
We assume that the channel statistics of all IoT devices are the same, as given in Section \ref{sec:simu_single}.
For comparison, consider a zero-wait baseline policy, i.e., in each slot, if an IoT device is scheduled to update its status packet, then it will sample the physical process immediately, \textcolor{black}{which takes one slot}.
\textcolor{black}{For the zero-wait baseline policy, the updating control and the updates of the per-device Q-factors and the Lagrange multipliers are similar to those of the proposed suboptimal policy, i.e., Step~\ref{step:auction} and Step~\ref{step:update} in Algorithm~\ref{alg:learning}.}
  This is a commonly used baseline in the literature on AoI minimization, e.g., see \cite{8000687} and references therein.

\textcolor{black}{In Fig.~\ref{fig:performance_op_sub}, we compare the average AoI at the destination, resulting from the optimal policy, the proposed semi-randomized policy, and the zero-wait baseline policy, for two IoT devices under different values of $C^{\textrm{max}}_k$. From Fig.~\ref{fig:performance_op_sub}, we can see that the proposed semi-distributed policy achieves a near-optimal performance and significantly outperforms the zero-wait policy.}

Fig.~\ref{fig:aoi_cost_numusers} shows the average, per-device AoI at the destination and the average, per-device energy cost, resulting from the proposed semi-distributed policy and the  zero-wait baseline policy. The simulation results are obtained by averaging over 100,000 time slots.
\textcolor{black}{In the simulations, for $K=10,20,30,40,50$, it takes about $17000, 23000, 30000, 41000, 50000$ time slots for the convergence of the proposed suboptimal policy, respectively.}
% This is due to that the size of state space grows exponentially with $K$.
From Fig.~\ref{fig:aoi_cost_numusers}, we can see that the proposed semi-distributed policy can achieve up-to 20\% reduction of the average AoI at the destination  over the zero-wait baseline policy, with similar energy costs. Thus the proposed policy can make better use of the limited energy at the IoT device. Moreover, for both policies, we observe that, as the number of the IoT devices increases, the average per-device AoI at the destination increases, while the average energy cost decreases. This is due to the fact that the transmission opportunities for each IoT device become lower with more IoT devices.

\begin{figure}[!t]
\begin{centering}
\includegraphics[scale=.6]{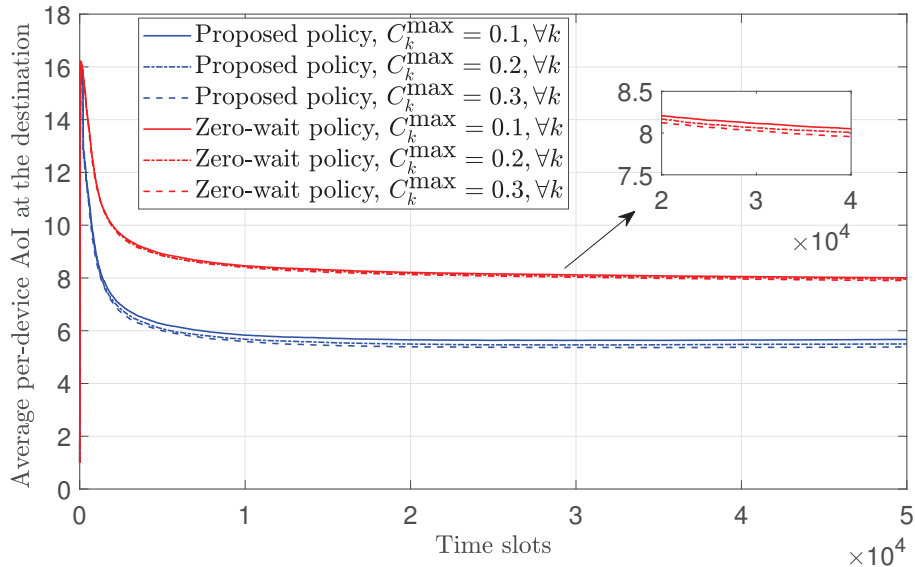}
% \vspace{-0.3cm}
 \caption{Illustration of the convergence property. The number of IoT devices is $K=5$. \textcolor{black}{$A_{l,k}^\textrm{max}=A_{r,k}^\textrm{max}=100, \forall k$.}}\label{fig:convergence}
\end{centering}
% \vspace{-0.7cm}
\end{figure}

In Fig.~\ref{fig:convergence}, we \textcolor{black}{show} the evolution of the average per-device AoI at the destination, resulting from the proposed semi-distributed policy and the zero-wait baseline policy, under different  $C^{\textrm{max}}_k$. The convergence of the proposed semi-distributed learning algorithm can be clearly observed (after about 15,000 time slots). Moreover, with the increase of $C^{\textrm{max}}_k$, the average per-device AoI at the destination for the two policies decreases. The performance gain of the proposed policy over the baseline policy can be as much as $33\%$ when $C^{\textrm{max}}_k=0.3$.

% For the metric of the system-wide
% computation overhead, the distributed computation ofﬂoading
% solution can achieve up-to 68\% and 55\%, and 51\% overhead
% reduction over with the solutions by local computing by all
% users, and cloud computing by all users, respectively.

\section{Conclusion}
In this paper, we have studied the optimal sampling and updating processes that enable IoT devices to minimize the average AoI at the destination under an average energy constraint for each IoT device in a real-time IoT monitoring system.
We have formulated this problem as an infinite horizon average cost CMDP and transformed it into an unconstrained MDP.
For the single IoT device case, we have shown that the optimal sampling and updating policy is of threshold type, which reveals a fundamental tradeoff between the average AoI at the destination and the sampling and updating costs.
Based on this optimality property, we have proposed a structure-aware algorithm to obtain the optimal policy for the CMDP.
We  have also studied the effects of the wireless channel fading and shown that channels with large mean channel gain and less scattering can achieve better AoI performance.
For the case of multiple IoT devices, we have shown that the optimal sampling and updating policy is a function of the Q-factors of the unconstrained MDP.
To reduce the complexity in obtaining the optimal Q-factors, we have developed a semi-distributed low-complexity suboptimal policy by approximating the optimal Q-factors by a linear form of the per-device Q-factors. We have proposed an online algorithm for each device to estimate and learn its per-device Q-factor based on the locally observed AoI and channel states.
We have shown the almost surely convergence of the proposed  learning algorithm to the proposed suboptimal policy.
Simulation results have shown that, for the single IoT device case, the optimal thresholds for sampling (updating) are non-decreasing with the sampling (updating) cost and the optimal action is threshold-based with respect to the channel state; and the proposed semi-distributed suboptimal policy for multiple IoT devices yields significant performance gain in terms of the average AoI compared to a zero-wait baseline policy.
Future work will address key extensions such as  providing an approximation analysis of the considered linear decomposition method and proposing grant-free uplink transmission protocols.

\appendices
\section*{Appendix}

% \section{Proof of Lemma~\ref{lemma:bellman}}\label{app:bellman}

\subsection{Proof of Lemma~\ref{lemma:valuefunction}}\label{app:value_function}
We prove Lemma~\ref{lemma:valuefunction} using  the value iteration algorithm (VIA) and mathematical induction.
First, we introduce the VIA\cite[Chapter 4.3]{bertsekas}. For notational convenience, we omit $\lambda$ in the notation of $V(\bs{A},h;\lambda)$. For each state $(\bs{A},h)\in\mathcal{A}\times\mathcal{H}$, let $V_m(\bs{A},h)$ be the value function at iteration $m$.
Define the state-action cost function at iteration $m$ as:
\begin{align}
&J_{m+1}(\bs{A},h,\bs{w})\triangleq L(\bs{A},h,\bs{w};\lambda)+\sum_{h'\in\mathcal{H}}p_{\mathcal{H}}(h')V_m(\bs{A}',h'),\label{eqn:jl}
\end{align}
where $\bs{A}'$ is given by Lemma~\ref{lemma:bellman}.
Note that $J_{m+1}(\bs{A},h,\bs{w})$ is related to the right-hand side of the Bellman equation in \eqref{eqn:bellman}.
For each $(\bs{A},h)$, VIA calculates $V_{m+1}(\bs{A},h)$ according to
\begin{equation}\label{eqn:RVIA}
  V_{m+1}(\bs{A},h)=\min_{\bs{w}\in\mathcal{W}} J_{m+1}(\bs{A},h,\bs{w}),~\forall l.
\end{equation}
Under any initialization of $V_0(\bs{A},h)$, the generated sequence $\{V_m(\bs{A},h)\}$ converges to $V(\bs{A},h)$\cite[Proposition 4.3.1]{bertsekas}, i.e.,
 \begin{equation}
   \lim_{m\to\infty}V_m(\bs{A},h)=V(\bs{A},h),~\forall (\bs{A},h)\in\mathcal{A}\times\mathcal{H},\label{eqn:converge}
 \end{equation}
 where $V(\bs{A},h)$ satisfies the Bellman equation in \eqref{eqn:bellman}.
 Let $\pi^*_m(\bs{A},h)$  denote the control that attains the minimum of the first term in \eqref{eqn:RVIA} at iteration $m$ for all $\bs{A},h$, i.e.,
\begin{equation}
  \pi^*_m(\bs{A},h)=\arg\min_{\bs{w}\in\mathcal{W}} J_{m+1}(\bs{A},h,\bs{w}),~~\forall (\bs{A},h)\in\mathcal{A}\times\mathcal{H}.\label{eqn:optimal_l}
\end{equation}
We refer to $\pi^*_m=(\pi^*_{s,m},\pi^*_{u,m})$ as the optimal policy for iteration $m$.

Now, consider two AoI states, $\bs{A}^1=(A_l^1,A_r^1)$ and $\bs{A}^2 = (A_l^2,A_r^2)$.
To prove Lemma~\ref{lemma:valuefunction}, we only need to show that for any $\bs{A}^1,\bs{A}^2\in\mathcal{A}$, such that $A_l^2\geq A_l^1$ and $A_r^2\geq A_r^1$,
\begin{equation}
V_{m}(\bs{A}^2,h)\geq V_{m}(\bs{A}^1,h),\label{eqn:vl}
\end{equation}
holds for all $m=0,1,\cdots$.

First, we initialize $V_{0}(\bs{A},h)=0$ for all $\bs{A},h$. Thus, \eqref{eqn:vl} holds for $m=0$.
Assume that \eqref{eqn:vl} holds for  some $m>0$. We will prove that \eqref{eqn:vl} also holds for $m+1$. By \eqref{eqn:RVIA},
we have
\begin{align}
V_{m+1}(\bs{A}^1,h)&=J_{m+1}\left(\bs{A}^1,h,\pi^*_m(\bs{A}^1,h)\right)
\overset{(a)}{\leq}J_{m+1}\left(\bs{A}^1,h,\pi^*_m(\bs{A}^2,h)\right)\nonumber\\
&\overset{(b)}{=} A_r^1 + \lambda C(\pi^*_m(\bs{A}^2,h)) + \sum_{h'\in\mathcal{H}}p_{\mathcal{H}}(h')V_m(A_l^{1'},A_r^{1'},h'),
\label{eqn:vl1q1}
\end{align}
where  $(a)$ is due to the optimality of $ \pi^*_m(\bs{A}^1,h)$ for $(\bs{A}^1,h)$ in the $m$-th iteration, $(b)$ directly follows from \eqref{eqn:jl}, $A_l^{1'} = \min\{\pi^*_{s,m}(\bs{A}^2,h)+(1-\pi^*_{s,m}(\bs{A}^2,h))(A_l^1+1),\hat{A}_l\}$
and $A_r^{1'} =\min\{\pi^*_{u,m}(\bs{A}^2,h)(A_l^1+1)+(1-\pi^*_{u,m}(\bs{A}^2,h))(A_r^1+1),\hat{A}_r\}$. By \eqref{eqn:jl} and \eqref{eqn:RVIA}, we also have
\begin{align*}
V_{m+1}(\bs{A}^2,h)=J_{m+1}\left(\bs{A}^2,h,\pi^*_m(\bs{A}^2,h)\right)
=A_r^2 + \lambda C(\pi^*_m(\bs{A}^2,h)) + \sum_{h'\in\mathcal{H}}p_{\mathcal{H}}(h')V_m(A_l^{2'},A_r^{2'},h'),
% \label{eqn:vl1q2}
\end{align*}
where $A_l^{2'} = \min\{\pi^*_{s,m}(\bs{A}^2,h)+(1-\pi^*_{s,m}(\bs{A}^2,h))(A_l^2+1),\hat{A}_l\}$
and $A_r^{2'} =\min\{\pi^*_{u,m}(\bs{A}^2,h)(A_l^2+1)+(1-\pi^*_{u,m}(\bs{A}^2,h))(A_r^2+1),\hat{A}_r\}$.

It can be seen that $A_l^{2'}\geq A_l^{1'}$ and $A_r^{2'}\geq A_r^{1'}$ for all possible $\pi^*_m(\bs{A}^2,h)\in\mathcal{W}$, which implies that $V_m(A_l^{2'},A_r^{2'},h')\geq V_m(A_l^{1'},A_r^{1'},h')$ by induction. Thus, we have $V_{m+1}(\bs{A}^2,h)\geq V_{m+1}(\bs{A}^1,h)$, i.e., \eqref{eqn:vl} holds for $m+1$. Therefore, by induction, we can show that \eqref{eqn:vl} holds for any $m$.
 By taking limits on both sides of \eqref{eqn:vl} and by \eqref{eqn:converge}, we complete the proof of Lemma~\ref{lemma:valuefunction}.

\subsection{Proof of Lemma~\ref{lemma:stateactionfunction} }\label{app:state_action_function}
First, we derive the general relation between $\Delta J_{\bs{w},\bs{w}'}(\bs{A}^1,h)$ and $\Delta J_{\bs{w},\bs{w}'}(\bs{A}^2,h)$ for any $\bs{w},\bs{w}'\in\mathcal{W}$, $h\in\mathcal{H}$, and $\bs{A}^1,\bs{A}^2\in\mathcal{A}$. Here,  $\lambda$ is also omitted in the notation of $\Delta J_{\bs{w},\bs{w}'}(\bs{A}^1,h;\lambda)$ for notational convenience. By \eqref{eqn:state-action-function}, we have
\begin{align}
&\Delta J_{\bs{w},\bs{w}'}(\bs{A}^1,h) - \Delta J_{\bs{w},\bs{w}'}(\bs{A}^2,h)
\nonumber\\
&= \Big(L(\bs{A}^1,h,\bs{w})+\sum_{h'\in\mathcal{H}}p_{\mathcal{H}}(h')V(\bs{A}^{1,\bs{w}},h') - L(\bs{A}^1,h,\bs{w}')+\sum_{h'\in\mathcal{H}}p_{\mathcal{H}}(h')V(\bs{A}^{1,\bs{w}'},h')\Big)\nonumber\\
&~- \Big(L(\bs{A}^2,h,\bs{w})+\sum_{h'\in\mathcal{H}}p_{\mathcal{H}}(h')V(\bs{A}^{2,\bs{w}},h') - L(\bs{A}^2,h,\bs{w}')+\sum_{h'\in\mathcal{H}}p_{\mathcal{H}}(h')V(\bs{A}^{2,\bs{w}'},h')\Big)\nonumber\\
&= \sum_{h'\in\mathcal{H}}p_{\mathcal{H}}(h')\left(V(\bs{A}^{1,\bs{w}},h')-V(\bs{A}^{1,\bs{w}'},h')-V(\bs{A}^{2,\bs{w}},h')+V(\bs{A}^{2,\bs{w}'},h')\right),
\label{eqn:delta_delta_J}
\end{align}
where $A_l^{1,\bs{w}} = \min\{s+(1-s)(A_l^1+1),\hat{A}_l\}$, $A_r^{1,\bs{w}} =\min\{u(A_l^1+1)+(1-u)(A_r^1+1),\hat{A}_r\}$, $A_l^{1,\bs{w}'} = \min\{s'+(1-s')(A_l^1+1),\hat{A}_l\}$, $A_r^{1,\bs{w}'} =\min\{u'(A_l^1+1)+(1-u')(A_r^1+1),\hat{A}_r\}$, $A_l^{2,\bs{w}} = \min\{s+(1-s)(A_l^2+1),\hat{A}_l\}$, $A_r^{2,\bs{w}} =\min\{u(A_l^2+1)+(1-u)(A_r^2+1),\hat{A}_r\}$, $A_l^{2,\bs{w}'} = \min\{s'+(1-s')(A_l^2{}+1),\hat{A}_l\}$, and $A_r^{2,\bs{w}'} =\min\{u'(A_l^2+1)+(1-u')(A_r^2+1),\hat{A}_r\}$.

Next, based on \eqref{eqn:delta_delta_J}, we \textcolor{black}{show that $\Delta J_{\bs{w},\bs{w}'}(\bs{A}^1,h)$ is non-decreasing with $A_l$ for $\bs{w}'=(1,0)$}. Consider $\bs{w}=(0,0)$, $\bs{w}'=(1,0)$, and $\bs{A}^1$ and $\bs{A}^2$ satisfying \textcolor{black}{$A_l^1\geq A_1^2$} and $A_r^1=A_r^2$. We can see that,  $A_l^{1,\bs{w}}\geq A_l^{2,\bs{w}}$, $A_r^{1,\bs{w}}=A_r^{2,\bs{w}}$, $A_l^{1,\bs{w}'}=A_l^{2,\bs{w}'}$, and \textcolor{black}{$A_r^{1,\bs{w}'}=A_r^{2,\bs{w}'}$}. Thus, we have $V(\bs{A}^{1,\bs{w}'},h')= V(\bs{A}^{2,\bs{w}'},h')$ and  by Lemma~\ref{lemma:valuefunction}, we have $V(\bs{A}^{1,\bs{w}},h')\geq V(\bs{A}^{2,\bs{w}},h')$. Therefore, by \eqref{eqn:delta_delta_J}, we have $\Delta J_{\bs{w},\bs{w}'}(\bs{A}^1,h) - \Delta J_{\bs{w},\bs{w}'}(\bs{A}^2,h)\geq 0$, which completes the proof. Similarly, we can also \textcolor{black}{prove the remaining properties of $\Delta J_{\bs{w},\bs{w}'}(\bs{A}^1,h)$ in Lemma~\ref{lemma:stateactionfunction}.}

\subsection{Proof of Theorem~\ref{theorem:optimal}}\label{app:thereom1}

We first prove Property A) of Theorem~\ref{theorem:optimal}. Consider action $\bs{w}=(0,0)$, action $\bs{w}'=(1,0)$, channel state $h$, AoI state $\bs{A}$ where $A_l=\phi_{(0,0)}^+(A_r,h)$. ($\lambda$ is omitted here.) Note that, we only need to consider \textcolor{black}{$\phi_{(0,0)}^+(A_r,h)>-\infty$}. According to the definition of $\phi_{\bs{w}}^+(A_r,h)$ in \eqref{eqn:func1}, we can see that $\Delta J_{\bs{w},\bs{w}'}(\bs{A},h)\leq 0$, i.e., $\bs{w}$ dominates $\bs{w}'$ for state $(\bs{A},h)$. Now, consider another AoI state $\bs{A}'$ where $A_l'\leq A_l$ and $A_r'=A_r$. By Lemma~\ref{lemma:stateactionfunction}, we obtain that
 \begin{equation}
 \Delta J_{\bs{w},\bs{w}'}(\bs{A}',h) \leq \Delta J_{\bs{w},\bs{w}'}(\bs{A},h)\leq 0,\label{eqn:proof_theorem1_eqn1}
\end{equation}
i.e., $\bs{w}=(0,0)$ also dominates $\bs{w}'=(1,0)$ for state $(\bs{A}',h)$.
Now, we consider $\bs{w}'=(0,1)$ or $(1,1)$,  channel state $h$, AoI state $\bs{A}$ where $A_l=\psi_{(0,0)}^+(A_l,h)$, AoI state $\bs{A}'$ where $A_l'= A_l$ and $A_r'\leq A_r$. According to the definition of $\psi_{\bs{w}}^+(A_l,h)$ in \eqref{eqn:func3} and Lemma~\ref{lemma:stateactionfunction}, we can prove that \eqref{eqn:proof_theorem1_eqn1} still holds, i.e., $\bs{w}=(0,0)$ also dominates $\bs{w}'=(1,0)$ or $(1,1)$ for state $(\bs{A}',h)$. By the definition of $\mathcal{A}_0(h)$, we can see that if $\bs{A}\in\mathcal{A}_0(h)$, then $\bs{w}=(0,0)$ dominates all other actions, i.e., $\pi^*(\bs{A},h)=(0,0)$. We complete the proof of Property A).

Next, we prove Property B) of Theorem~\ref{theorem:optimal}. Consider action $\bs{w}=(0,1)$, channel state $h$, AoI state $\bs{A}$ where $A_r=\psi_{(0,1)}^-(A_l,h)$. We only need consider that $\psi_{(0,1)}^-(A_l,h)<+\infty$. By the definition of $\psi_{(0,1)}^-(A_l,h)$ in \eqref{eqn:func4}, we have $\Delta J_{\bs{w},\bs{w}'}(\bs{A},h)\leq 0$ for all $\bs{w}'\neq\bs{w}$, i.e., $\pi^*(\bs{A},h)=(0,1)$. Now consider another AoI state $\bs{A}'$ where $A_l'= A_l$ and $A_r'\geq A_r$. By Lemma~\ref{lemma:stateactionfunction}, we can see that $\Delta J_{\bs{w},\bs{w}'}(\bs{A}',h) \leq \Delta J_{\bs{w},\bs{w}'}(\bs{A},h)\leq 0$ holds for all $\bs{w}'\neq\bs{w}$, i.e., $\pi^*(\bs{A}',h)=(0,1)$. We complete the proof of Property B). Following the proof of Property B), we can also prove Properties C) and D).  This completes the proof of Theorem~\ref{theorem:optimal}.

\subsection{Proof of Theorem~\ref{theorem:first_order}}\label{app:thereom2}

To prove Theorem~\ref{theorem:first_order}, we first prove that for any channels $I$ and $J$ such that $h^I$ first-order stochastically dominates $h^J$,
\begin{equation}
V^I(\bs{A},h;\lambda)\leq V^J(\bs{A},h;\lambda), \label{eqn:V_IJ}
\end{equation}
holds for all $(\bs{A},h)$,
where $V^I(\bs{A},h;\lambda)$ and $V^J(\bs{A},h;\lambda)$ are the value functions under channels $I$ and $J$, respectively.
We prove \eqref{eqn:V_IJ} through mathematical induction and the VIA in Appendix A.
Similar to Appendix A, we introduce $V_m^I(\bs{A},h;\lambda)$,$V_m^J(\bs{A},h;\lambda)$, $J_{m}^I(\bs{A},h,\bs{w})$, $J_{m}^J(\bs{A},h,\bs{w})$,  $\pi^{I*}_m=(\pi^{I*}_{s,m},\pi^{I*}_{u,m})$, and $\pi^{J*}_m=(\pi^{J*}_{s,m},\pi^{J*}_{u,m})$ for channels $I$ and $J$.
Since $C_u(h)$ is non-increasing with $h$, it can be easily shown that $V_m^I(\bs{A},h;\lambda)$ and $V_m^J(\bs{A},h;\lambda)$ are non-increasing with $h$, by using induction and the VIA.
To show \eqref{eqn:V_IJ}, by \eqref{eqn:converge}, we only need to show that
\begin{equation}
V^I_m(\bs{A},h;\lambda)\leq V^J_m(\bs{A},h;\lambda), \label{eqn:V_IJ_m}
\end{equation}
holds for $m=0,1,\cdots$. We initialize $V^I_0(\bs{A},h;\lambda)= V^J_0(\bs{A},h;\lambda)=0$, for all $(\bs{A},h)$, i.e., \eqref{eqn:V_IJ_m} holds for $m=0$.
Assume that \eqref{eqn:V_IJ_m} holds for some $m>0$. We will show that \eqref{eqn:V_IJ_m} also holds for $m+1$.
By  \eqref{eqn:jl} and \eqref{eqn:RVIA}, we have
\begin{align*}
V_{m+1}^I(\bs{A},h;\lambda)&=J_{m+1}^I\left(\bs{A},h,\pi^{I*}_m(\bs{A},h);\lambda\right)\overset{(c)}{\leq}J_{m+1}^I\left(\bs{A},h,\pi^{J*}_m(\bs{A},h);\lambda\right)\nonumber\\
&= L(\bs{A},h,\pi^{J*}_m(\bs{A},h);\lambda) + \sum_{h'\in\mathcal{H}}p_{\mathcal{H}}^I(h')V_m^I(\bs{A}',h';\lambda) \nonumber\\
&\overset{(d)}{\leq} L(\bs{A},h,\pi^{J*}_m(\bs{A},h);\lambda) + \sum_{h'\in\mathcal{H}}p_{\mathcal{H}}^J(h')V_m^I(\bs{A}',h';\lambda)\nonumber\\
&\overset{(e)}{\leq} L(\bs{A},h,\pi^{J*}_m(\bs{A},h);\lambda) + \sum_{h'\in\mathcal{H}}p_{\mathcal{H}}^J(h')V_m^J(\bs{A}',h';\lambda)
=V_{m+1}^J(\bs{A},h;\lambda),
\end{align*}
where  $(c)$ is due to the optimality of $ \pi^{I*}_m(\bs{A},h)$ for $(\bs{A},h)$ under channel $I$ in the $m$-th iteration,
$(d)$ is due to that  $h^I$ first-order stochastically dominates $h^J$ and $V_m^I(\bs{A},h;\lambda)$ is non-increasing with $h$,
$(e)$ follows from the induction hypothesis $V^I_m(\bs{A},h;\lambda)\leq V^J_m(\bs{A},h;\lambda)$,
$A_l' = \min\{\pi^{J*}_{s,m}(\bs{A},h)+(1-\pi^{J*}_{s,m}(\bs{A},h))(A_l+1),\hat{A}_l\}$,
and $A_r' =\min\{\pi^{J*}_{u,m}(\bs{A},h)(A_l+1)+(1-\pi^{J*}_{u,m}(\bs{A},h))(A_r+1),\hat{A}_r\}$.
Thus, we prove \eqref{eqn:V_IJ_m} holds for $m+1$. Then, by induction and \eqref{eqn:converge}, we can show that \eqref{eqn:V_IJ} holds.
Based on \eqref{eqn:V_IJ}, Lemma~\ref{lemma:bellman} and Propositions 4.3.1 in\cite{bertsekas}, we have
% \begin{equation}
$\bar{L}^{I*}(\lambda) = \theta_{\lambda}^I \leq \theta_{\lambda}^J = \bar{L}^{J*}(\lambda). $
% \end{equation}
Finally, by Lemma~\ref{lemma:relation}, we can see that
% \begin{equation}
$\bar{A}_r^{I*} = \max_{\lambda\geq 0} \bar{L}^{I*}(\lambda) - \lambda C^{\textrm{max}} \leq \max_{\lambda\geq 0} \bar{L}^{J*}(\lambda) - \lambda C^{\textrm{max}} =\bar{A}_r^{J*},$
% \end{equation}
which completes the proof of Theorem~\ref{theorem:first_order}.

\subsection{Proof of Lemma~\ref{lemma:bellman_miot}}\label{app:lemma5}
For a given $\bs\lambda$, by Propositions 4.2.1, 4.2.3, and 4.2.5 in \cite{bertsekas} \textcolor{black}{(see Propositions 4.2.3 and 4.2.5 in Appendix H)}, the optimal average Lagrange cost for the unconstrained MDP in \eqref{eqn:opt_L_miot} is the same for all initial states and the optimal policy  can be obtained by solving the following Bellman equation with respect to $(\theta_{\bs\lambda},\{V(\bs{A},\bs{h};\bs\lambda)\})$.
\begin{align*}
  \theta_{\bs\lambda}+V(\bs{A},\bs{h};\bs\lambda)=\min_{\bs{w}\in\mathcal{W}}\Bigg\{L(\bs{A},\bs{h},\bs{w};\bs\lambda)+\sum_{\bs{h}'\in\mathcal{H}}p_{\mathcal{H}}(\bs{h}') V(\bs{A}',\bs{h}';\bs\lambda)\Bigg\},
  ~\forall (\bs{A},\bs{h})\in\mathcal{A}\times\mathcal{H},
\end{align*}
where $V(\bs{A},\bs{h};\bs\lambda)$ is the value function. Since $\pi^*_{\lambda}(\bs{A},\bs{h})= (\pi_{\bs\lambda,s}^*(\bs{A},\bs{h}),\pi_{\bs\lambda,u}^*(\bs{A},\bs{h}))$, we introduce the Q-factor of state $(\bs{A},\bs{h})$ under updating action $\bs{u}$ as:
 \begin{equation*}
 Q(\bs{A},\bs{h},\bs{u};\bs\lambda)\triangleq\min_{\bs{s}\in\mathcal{S}}\{L(\bs{A},\bs{h},\bs{w};\bs\lambda)+\sum_{\bs{h}'\in\mathcal{H}}p_{\mathcal{H}}(\bs{h}') V(\bs{A}',\bs{h}';\bs\lambda)\}-\theta_{\bs\lambda}.
 \end{equation*}
    Thus, we have $V(\bs{A},\bs{h};\bs\lambda) = \min_{\bs{u}\in\mathcal{U}}Q(\bs{A},\bs{h},\bs{u};\bs\lambda)$ for all $(\bs{A},\bs{h})$ and  $(\theta_{\bs\lambda},\{Q(\bs{A},\bs{h},\bs{u};\bs\lambda)\})$ satisfies the Bellman equation in \eqref{eqn:bellman_miot}. We complete the proof.
    % of Lemma~\ref{lemma:bellman_miot}.

\subsection{Proof of Lemma~\ref{lemma:convergence_of_Q_factor}}\label{app:lemma6}
Under a unichain policy defined in Definition~\ref{definition:stationary_policy_MIoT}, the induced random process $\{(\bs{A}(t),\bs{h}(t))\}$ is a controlled Markov chain with a single recurrent class and possibly some transient states \cite{bertsekas}.
According to the explanation for the condition of Proposition 4.3.2 in \cite{bertsekas}, the condition of Lemma 2 in \cite{5462936} is satisfied for our problem. Then, by following the proofs of Lemma 2 in \cite{5462936} and Proposition 4.3.2 in \cite{bertsekas}, we can prove the Lemma~\ref{lemma:convergence_of_Q_factor}. The detailed proof is omitted due to page limitations.

\subsection{Proof of Lemma~\ref{lemma:convergence_of_lambda}}\label{app:lemma7}
Due to the separation of the two timescales of the updates in \eqref{eqn:update_Q_factor} and \eqref{eqn:update_lambda}, the update of the Q-factors can be regarded as converged to $\bs{Q}_k^{\infty}(\bs{\lambda}^t)$ under $\bs{\lambda}^t$\cite{stochasticlearning}. Then, by the theory of stochastic approximation \cite{5462936,stochasticlearning,BORKAR1997291}, the iterations of the update of the Lagrange multiplier in \eqref{eqn:update_lambda} can be described by the following Ordinary Differential Equation (ODE):
\begin{align}\label{eqn:ode}
\dot{\bs{\lambda}^t} = \mathbb{E}^{\pi_{\bs{\lambda}^t}^*}[C_1(\hat{\bs{w}}_1(t))-C_1^{\textrm{max}},\cdots,C_K(\hat{\bs{w}}_K(t))-C_K^{\textrm{max}}],
\end{align}
where $\pi_{\bs{\lambda}^t}^*$ is the converged control policy in Algorithm~\ref{alg:learning} under $\bs{\lambda}^t$ and the expectation is taken with respect to the measure induced by the policy $\pi_{\bs{\lambda}^t}^*$.
Denote $\bar{L}(\bs\lambda^t)=\mathbb{E}^{\pi_{\bs{\lambda}^t}^*}[\sum_{k=1}^K (A_{r,k}  + \allowbreak\lambda_k (C_k(\hat{\bs{w}}_k)-C^{\textrm{max}}_k))]$.
Since the sampling and updating actions are discrete, we have $\pi_{\bs{\lambda}^t}^*=\pi_{\bs{\lambda}^t+\bs\delta_{\lambda}}^*$.
By chain rule, it can be seen that $\frac{\partial{\bar{L}(\bs\lambda^t)}}{\partial{\lambda^t_k}}=\mathbb{E}^{\pi_{\bs{\lambda}^t}^*}[C_k(\hat{\bs{w}}_k(t))-C_k^{\textrm{max}}]$.
Thus, the ODE in \eqref{eqn:ode} can be expressed as $\dot{\bs{\lambda}^t}=\triangledown \bar{L}(\bs\lambda^t)$. Therefore, the ODE in \eqref{eqn:ode} will converge to $\arg\max\bar{L}(\bs\lambda^{\infty})$, which corresponds to $\triangledown \bar{L}(\bs\lambda^{\infty})=0$. In other words, the policy under $(\bs{Q}^{\infty},\bs\lambda^{\infty})$ satisfies the constraint in \eqref{eqn:constraint_miot}. This completes the proof.

\vspace{-1cm}
\textcolor{black}{
\subsection{Some preliminaries on MDP}\label{app:pre}
Proposition 4.2.3 in \cite{bertsekas}: Let the weak accessibility (WA) condition hold. Then the optimal average cost is the same for all initial states.}

\textcolor{black}{Proposition 4.2.5 in \cite{bertsekas}: If all stationary policies are unichain, the WA condition holds.}

\textcolor{black}{Theorem 8.6.6 in \cite{puterman}: Suppose all stationary policies are unichain, and the set of states and actions are finite, then policy iteration converges in a finite number of iterations to the optimal policy satisfying the Bellman equation.
}

\bibliographystyle{IEEEtran}
\bibliography{IEEEabrv,AoI}

\end{document}